\documentstyle[aps,eqsecnum,preprint]{revtex}

\def\pmb#1{\leavevmode\setbox0=\hbox{$#1$}
\kern-.025em\copy0\kern-\wd0
\kern-.05em\copy0\kern-\wd0\kern-.025em\raise.0433em\box0}

\catcode`@=11
\def\lsim{\mathchoice
  {\mathrel{\lower.8ex\hbox{$\displaystyle\buildrel<\over\sim$}}}
  {\mathrel{\lower.8ex\hbox{$\textstyle\buildrel<\over\sim$}}}
  {\mathrel{\lower.8ex\hbox{$\scriptstyle\buildrel<\over\sim$}}}
  {\mathrel{\lower.8ex\hbox{$\scriptscriptstyle\buildrel<\over\sim$}}} }
\def\gsim{\mathchoice
  {\mathrel{\lower.8ex\hbox{$\displaystyle\buildrel>\over\sim$}}}
  {\mathrel{\lower.8ex\hbox{$\textstyle\buildrel>\over\sim$}}}
  {\mathrel{\lower.8ex\hbox{$\scriptstyle\buildrel>\over\sim$}}}
  {\mathrel{\lower.8ex\hbox{$\scriptscriptstyle\buildrel>\over\sim$}}} }
\def\croce{\displaystyle / \kern-0.2truecm\hbox{$\backslash$}}
\def\lqua{\lower4pt\hbox{\kern5pt\hbox{$\sim$}}\raise1pt
\hbox{\kern-8pt\hbox{$<$}}~}
\def\gqua{\lower4pt\hbox{\kern5pt\hbox{$\sim$}}\raise1pt
\hbox{\kern-8pt\hbox{$>$}}~}
\def\mma{\lower1pt\hbox{\kern5pt\hbox{$\scriptstyle <$}}\raise2pt
\hbox{\kern-7pt\hbox{$\scriptstyle >$}}~}
\def\mmb{\lower1pt\hbox{\kern5pt\hbox{$\scriptstyle >$}}\raise2pt
\hbox{\kern-7pt\hbox{$\scriptstyle <$}}~}
\def\mmc{\lower4pt\hbox{\kern5pt\hbox{$<$}}\raise1pt
\hbox{\kern-8pt\hbox{$>$}}~}
\def\mmd{\lower4pt\hbox{\kern5pt\hbox{$>$}}\raise1pt
\hbox{\kern-8pt\hbox{$<$}}~}
\def\lsu{\raise4pt\hbox{\kern5pt\hbox{$\sim$}}\lower1pt
\hbox{\kern-8pt\hbox{$<$}}~}
\def\gsu{\raise4pt\hbox{\kern5pt\hbox{$\sim$}}\lower1pt
\hbox{\kern-8pt\hbox{$>$}}~}
\def\croce{\displaystyle / \kern-0.2truecm\hbox{$\backslash$}}
\def\ali{\hbox{A \kern-.9em\raise1.7ex\hbox{$\scriptstyle \circ$}}}
\def\2frecce{\hbox{\lower 0.3ex\hbox{$\leftarrow$}
\hbox{\kern-1.3em\raise 0.3ex\hbox{$\rightarrow$}}}}
\def\quad@rato#1#2{{\vcenter{\vbox{
        \hrule height#2pt
        \hbox{\vrule width#2pt height#1pt \kern#1pt \vrule width#2pt}
        \hrule height#2pt} }}}
\def\quadratello{\mathchoice
\quad@rato5{.5}\quad@rato5{.5}\quad@rato{3.5}{.35}\quad@rato{2.5}{.25} }
\font\s@=cmss10\font\s@b=cmbx8
\def\reali{{\hbox{\s@ l\kern-.5mm R}}}
\def\m{{\hbox{\s@ l\kern-.5mm M}}}
\def\k{{\hbox{\s@ l\kern-.5mm K}}}
\def\naturali{{\hbox{\s@ l\kern-.5mm N}}}
\def\interi{{\mathchoice
 {\hbox{\s@ Z\kern-1.5mm Z}}
 {\hbox{\s@ Z\kern-1.5mm Z}}
 {\hbox{{\s@b Z\kern-1.2mm Z}}}
 {\hbox{{\s@b Z\kern-1.2mm Z}}}  }}
\def\complessi{{\hbox{\s@ C\kern-1.7mm\raise.4mm\hbox{\s@b l}\kern.8mm}}}
\def\toro{{\hbox{\s@ T\kern-1.9mm T}}}
\def\unity{{\hbox{\s@ 1\kern-.8mm l}}}
\font\bold@mit=cmmi10
\def\setbmit{\textfont1=\bold@mit}
\def\bmit#1{\hbox{\textfont1=\bold@mit$#1$}}

\def\Ai{\hbox{\hbox{${\cal A}$}}\kern-1.9mm{\hbox{${/}$}}}
\def\Vi{\hbox{\hbox{${\cal V}$}}\kern-1.9mm{\hbox{${/}$}}}
\def\Di{\hbox{\hbox{${\cal D}$}}\kern-1.9mm{\hbox{${/}$}}}
\def\lam{\hbox{\hbox{${\lambda}$}}\kern-1.6mm{\hbox{${/}$}}}
\def\D{\hbox{\hbox{${D}$}}\kern-1.9mm{\hbox{${/}$}}}
\def\A{\hbox{\hbox{${A}$}}\kern-1.8mm{\hbox{${/}$}}}
\def\V{\hbox{\hbox{${V}$}}\kern-1.9mm{\hbox{${/}$}}}
\def\parz{\hbox{\hbox{${\partial}$}}\kern-1.7mm{\hbox{${/}$}}}
\def\B{\hbox{\hbox{${B}$}}\kern-1.7mm{\hbox{${/}$}}}
\def\R{\hbox{\hbox{${R}$}}\kern-1.7mm{\hbox{${/}$}}}
\def\si{\hbox{\hbox{${\xi}$}}\kern-1.7mm{\hbox{${/}$}}}
\def\be{\begin{equation}}
\def\ee{\end{equation}}
\def\ba{\begin{array}}
\def\ea{\end{array}}
\def\bea{\begin{eqnarray}}
\def\eea{\end{eqnarray}}
\def\dsp{\displaystyle}
\def\tst{\textstyle}

\begin{document}

\draft
\tighten

\title{$U(1)\times SU(2)$ Chern--Simons gauge theory of\\
 underdoped cuprate superconductors}
\author{P.A. Marchetti}
\address{  Dipartimento di Fisica ``G. Galilei'',
INFN, U. of Padova, I--35131 Padova, Italy}
\author{Zhao-Bin Su}
\address{Institute of Theoretical Physics, CAS, Beijing 100080, China}
\author{ Lu Yu}
\address{International Centre for Theoretical Physics, I-34100 Trieste, Italy\\
and Institute of Theoretical Physics, CAS, Beijing 100080, China}

\date{\today}
\maketitle

\begin{abstract}
The Chern-Simons bosonization  with $U(1)\times SU(2)$ gauge field
is applied to  the  2-D $t-J$ model  in the limit $t \gg J$,
to study the normal state properties of  underdoped
 cuprate superconductors.
We prove the
existence of an upper bound on the partition function for holons
in a  spinon background, and we find the  optimal spinon configuration
 saturating the upper bound on  average --
a coexisting flux phase and    $s+i d$-like RVB state.
 After neglecting
 the feedback of holon fluctuations on the
$U(1)$  field $B$ and spinon fluctuations on the $SU(2)$  field
$V$, the holon
field is  a fermion and the spinon field is a hard--core boson.
Within this approximation we show that the $B$ field  produces a
$\pi$ flux phase for the holons, converting them into Dirac--like fermions,
while the $V$ field, taking into account the feedback of holons
  produces a gap for the spinons vanishing in the
 zero doping limit. The  nonlinear $\sigma$-model with a mass term
describes the crossover from the short-ranged antiferromagnetic (AF)
state in doped samples to long range AF order in reference compounds.
Moreover,  we derive
a low--energy effective action in terms of spinons, holons and a
self-generated $U(1)$ gauge field. Neglecting  the gauge fluctuations,
the holons are described by the Fermi liquid theory with a Fermi
surface consisting of 4
``half-pockets'' centered at  $(\pm {\pi\over 2}, \pm {\pi \over
2})$ and  one  reproduces the results for the electron
spectral function obtained  in the mean field approximation, in
 agreement with the photoemission  data
 on underdoped cuprates. The gauge
fluctuations are not confining due to coupling to holons,
 but nevertheless yield an attractive
interaction between spinons and holons leading to a bound state
 with  electron quantum numbers.
The renormalisation effects due to gauge
fluctuations  give rise to non--Fermi liquid behaviour for the
composite  electron, in certain temperature range showing the linear
in $T$ resistivity.
This formalism provides a new interpretation of the spin gap in the
underdoped superconductors (mainly due to the short-ranged AF order)
and predicts that the minimal  gap for the physical electron is proportional
 to the square root of the doping concentration. Therefore
the gap does not vanish in any direction. All these predictions can be
checked explicitly in experiment.
\end{abstract}





\vfill\eject

\section{introduction}
\label{intro}

\subsection{Physical issue to be addressed}
\label{phys}

The proximity of superconductivity (SC) to antiferromagnetism (AF)
in reference compounds is a distinct feature of the high-$T_c$
superconductors.  Upon doping the AF goes away, giving rise to SC.
At the same time, the  Fermi surface (FS) is believed
to develop from small pockets around $(\pm {\pi \over 2}, \pm {\pi \over 2})$
\cite{shr}, anticipated for a doped Mott insulator,
  to a large one around $(\pi, \pi)$, expected from the electronic
structure calculations\cite{pic} and confirmed by
the angle-resolved photoemission spectroscopy (ARPES) experiments
\cite{shen}. To understand this crossover
is one of the key issues in resolving the high $T_c$ puzzle. For this reason,
the underdoped samples present particular interest due to the strong
interplay of SC with AF.

There is a consensus now that these systems are strongly
anisotropic, and the fundamental issue is to understand the behavior of
strongly correlated electrons in the copper-oxygen plane\cite{anbook}.
A ``spin gap'' or ``pseudogap'' has been invoked to explain\cite{mil} the
reduction of
magnetic susceptibility $\chi$ below certain characteristic temperature
$T^*$\cite{tak}
 and suppression of the specific heat compared with the linear
$T$ behavior\cite{lor}. This gap also shows up in
transport properties\cite{bat}, neutron scattering\cite{ros}, and NMR
relaxation
rate \cite{tak} measurements.
The recent ARPES experiments\cite{mar} on  underdoped
samples seem to indicate that the FS of these compounds is
probably half-pocket-like, {\it i.e.} a small pocket near $\Sigma$-point
$({\pi \over 2},
{\pi \over 2})$,
but lacking its outer part
in the reduced Brillioun zone scheme. These data show clear Fermi level
 crossing in the (0, 0) to $(\pi, \pi)$ direction, but no such crossing was
detected in the $(0, \pi)$ to $(\pi, \pi)$ direction\cite{mar}.
The observed pseudogap
above $T_c$ is consistent with d-wave symmetry.

In this paper we will be  concerned  with the normal
state properties of these underdoped cuprate superconductors,
focusing on the implications derived from the proximity of these systems
to the AF reference state.

\subsection{A brief survey of related theoretical approaches}

Theoretically there have been mainly two competing approaches: One starting
from the
Mott-Hubbard insulator, advocated by Anderson\cite{and,anbook} using the
concept of
spin liquid, or the Resonant Valence Bond (RVB) state,
 while  the other starting from the more conventional Fermi liquid (FL)
point of view.

One of the crucial concepts within the first approach is ``spin-charge
separation''
which can be intuitively implemented  by introducing
``slave bosons''\cite{zw}, namely, one rewrites the electron
operator:
$$\psi_{i\sigma} =e_i^{\dagger}f_{i\sigma},$$
where $e_i$ is a charged spinless
(slave) boson  operator (holon), while $f_{i\sigma}$ is a neutral, spin 1/2
fermion
operator (spinon) satisfying contraint
\begin{equation}
e_i^{\dagger}e_i + f^{\dagger}_{i\sigma}f_{i\sigma} =1.
\label{costr}\end{equation}
(Hereafter the repeated spin indices are summed over.)
One can also interchange the role of boson and fermion operators, i.e.,
to introduce a spinless fermion to describe the charge degree of
freedom, while ``spinning'' bosons to describe the spin degree of freedom.
This is the ``slave fermion'' approach\cite{yos,aro}. The essential
requirement for both approaches is the ``single occupancy''
 constraint which is very difficult to
implement. In the mean field approximation (MFA) which satisfies the
constraint (\ref{costr}) only on average, the ``slave boson''\cite{fuk} and
``slave fermion''
\cite{aro,ka} approaches gave very different phase diagrams and each of
them has its own difficulties\cite{yu}.  There have been several attempts to
improve the situation\cite{feng}, but the basic difficulty still remains.

Moreover, in decomposing the physical electron into a product of fermion
and boson, one increases the number of degrees of freedom (d.o.f.) by two.
The constraint (\ref{costr}) takes care of one, but there is one extra d.o.f.
 which corresponds to the spinon-holon gauge field. In fact,
the physical electron operator is invariant under the transformation:
$$ e_j \rightarrow e_j e^{i\xi_j}, \;\;\; f_{j\sigma} \rightarrow
f_{j\sigma} e^{i\xi_j},
$$
so one can ``gauge-fix'' $\xi_j$ according to the choice.
This is the starting point of the gauge field approach to the
strongly correlated electron systems\cite{ba,il}. This approach has been
systematically pursued by P.A. Lee and his collaborators first
as the $U(1)$ gauge theory\cite{ln}, and recently by considering the
$SU(2)$ gauge group\cite{wen}.

From the FL point of view the interplay of AF with SC, and the evolvement
of the FS with doping has been elaborated by Kampt and
Schrieffer\cite{ks}, and by Chubukov
and his collaborators\cite{chu}.  Very recently, Zhang\cite{zha}
has proposed an interesting $SO(5)$ model to consider the
AF-SC interplay.

\subsection{Basic Idea of Chern-Simons Bosonization}
\label{C-S basic}

In this paper we will use the C-S bosonization as the basic technical tool.
The procedure of reformulating the fermion problem in terms of bosons
 was pioneered by
the Jordan-Wigner transformation\cite{jw}
\begin{equation}
c^\dagger_j = a^\dagger_j e^{\tst -i\pi {\sum_{l<j}} a^\dagger_la_l},
\label{jw}
\end{equation}
where $c_j$ is a fermion operator, while $a_j$ is a hard-core
boson operator on a linear chain. The abelian bosonization
for one-dimensional fermions with linear dispersion\cite{lm}
is similar in spirit and has found extensive applications
in condensed matter physics\cite{fra,gnt}. The key formula there
is the Mandelstam representation (in field-theoretical jargons)\cite{lm,man}:
\begin{equation}
\psi_\alpha (x) \sim e^{\tst {i \over \sqrt{\pi}}
{ \int^x_{-\infty}} dy \dot{\phi} (y)
- (-1)^\alpha i\sqrt{\pi}\phi (x)},
\label{mand}\end{equation}
where $\psi$ is the fermion operator,
while $\phi$ is the boson operator, and $\alpha = 1, 2$. (Here we
have omitted the normal ordering of the operators.) A
rigorous derivation of (\ref{mand}) in terms of path integrals was given
in Ref. \cite{fm88}. This bosonization procedure was generalized by Witten
to the non-abelian  case\cite{wit}, which  reformulates the 1+1 dimensional
relativistic fermion problem with symmetry group $G$  as
$G$-valued nonlinear $\sigma$ (NL$\sigma$) model.
 That scheme has been extensively
used  for the study of quantum spin systems\cite{aff86}.

The abelian bosonization procedure has been generalized to 2+1
dimensional systems\cite{sem}.  It is  in some sense
analogous to the Jordan-Wigner transformation.
The typical relation is given as:
\begin{equation}
\psi (x) \sim \phi (x) e^{ \tst i \int_{\gamma_x}
A_\mu (y) dy^\mu},
\label{2djw}\end{equation}
where $\psi (x)$ is the fermion field operator,
while $\phi (x)$ is the boson field operator, and
the two are related by the Chern-Simons (C-S) $U(1)$ gauge
field operator $A_\mu$. The integration in the exponent
is taken over an arbitrary path $\gamma_x$ in the
2D plane, running from $x$ to $\infty$.
The path integral will contain an extra factor $e^{-kS_{c.s.}}$ with the
C-S action
$$
S_{c.s.} =  { 1 \over {4\pi i}} \int d^3x
\epsilon^{\mu\nu\rho}A_\mu\partial_\nu A_\rho,
$$
and  the C-S  coefficient (level) $k= 1/ (2l+1),
l= 0,1,2...$. For non-relativistic fermions the
C-S constraint can be solved explicitly,
and the transformation (\ref{2djw}) becomes\cite{sem}:
\begin{equation}
\psi (x) \sim \phi (x) e^{\tst i(2l+1) \int d^2y
\Theta (x-y) \phi^\dagger(y)  \phi (y)},
\label{2djwa}\end{equation}
where $\Theta  (x-y) = \arctan{\frac{x^2-y^2}{x^1-y^1}}$.
An interesting application of this formula is the
analysis of the fractional quantum Hall effect at filling
$\nu = 1/ (2l+1)$ in terms of boson liquids\cite{zhk}.
The statistical transmutation, implemented by the abelian C-S gauge
field is a consequence of the Aharonov-Bohm effect, and it is limited to
abelian fractional statistics, characterized by the phase factor
$2\pi \theta$ with $\theta \in [0,1)$\cite{wil}.
In particular, $\theta = 0, { 1\over 2} $ corresponds to boson and fermion,
respectively, while $\theta = {1\over 4} $ corresponds to the ``semion'' case,
advocated by Laughlin, as a constituent quasiparticle in  high temperature
superconductors\cite{lau}.

The 2D analogue of the Jordan-Wigner formula (\ref{2djw}) and
(\ref{2djwa}) was originally derived as an operator identity\cite{sem}.
It was later on justified in the path integral form and generalized to the
non-abelian case\cite{fro}.
That paper was further extended to include the correlation
functions, elaborated for the case $G=U(1)\times SU(2)$
and applied to the $t-J$ model\cite{fm92} (see also \cite{mar93}).
Readers are referred to those references for a detailed presentation.
To make the present paper more self-contained, we briefly outline here
the basic idea of such a bosonization procedure.

	Consider a system of N spin $ { 1 \over 2}$ fermions ( or bosons )
in two space dimensions, in an external  (abelian) gauge field $A$.
The canonical partition function in the first quantized path integral
representation is given by\cite{neg}:

\begin{equation}
Z(A) = {\displaystyle \sum_{\alpha_1,...\alpha_N} \int dx_1...dx_N
\sum_{\pi}
(-1)^{\sigma (\pi)}}
{\displaystyle \prod_{r=1}^N \Bigl(
\int_{\omega_r(o)=x_r}^{\omega_r(\beta)=x_{\pi(r)}}}
D\omega_r(\tau) e^{\tst  -{m \over 2}
\int \dot{\omega}_r^2 (\tau)}\Bigr)
e^{\tst  i\int_{\underline{\omega}}A},
\label{fk1}\end{equation}
where $\alpha_j = 1, 2$ for $j =1, 2, ....N$ are the spin indices,
$\omega_j (\tau)$
represent ``virtual trajectories'' of  particles
going from imaginary time $\tau=0$ to $\tau=\beta = 1/T$,
with $T$ as the temperature (we set the Boltzmann constant $k_B=1$)
 and reaching the plane $\tau=\beta$ at the same
set of points where they start at $\tau=0$, the points being arbitrarily
permuted. Due to the fermion statistics of the particles,
there is  a factor $(-1)^{\sigma(\pi)}$,
associated with each permutation $\pi$,
 where $\sigma (\pi)$ is the number of
exchanges in the permutation $\pi$. These trajectories have vanishing
probability to intersect each other for a given $\tau$.  Each set of such
trajectories appearing in the first quantized representation form a link,
i.e. a set of possibly interlaced loops, when the $\tau =0$ and $\tau
=\beta$ planes
are identified by periodicity in time.
Here   $\int_{\underline{\omega}} A = \int (A_l d \omega^l + A_0 d\tau)$ is
a line
integral in 2+1 dimensions, $ l=1,2.$
The partition function for the boson system would be the same,
except that $\sigma (\pi)$ is replaced with 0.  On the other hand,
  the factor $(-1)^{\sigma(\pi)}$ is  a topological
invariant naturally associated with the link and, according to a general
theory\cite{wit89}, it can be represented as the expectation value of a
``Wilson
loop'' (trace of a gauge phase factor) supported on that link in a gauge
theory with a suitably chosen C-S action.

$$(-1)^{\sigma(\pi)} = {\displaystyle\int DV e^{\tst -k S_{c.s.}(V)}
P(e^{\tst i\int_{\underline{\omega}} V })},$$
where $V$ is a C-S gauge field with symmetry group $G$, $k$ is the
C-S cofficient (level), already defined in the abelian case,
$P(\cdot)$ is  the path-ordering, identical to  the time-ordering,
if ``time'' is  parameterizing the path, and

\begin{equation}
S_{c.s.} = {\displaystyle{ 1 \over {4\pi i}} \int } d^3xTr[ \epsilon^
{\mu\nu\rho}(V_\mu
 \partial_\nu V_\rho + {2 \over 3} V_\mu  V_\nu  V_\rho)].
\label{cssu2}\end{equation}

As a consequence, there is a boson-fermion relation for the
canonical partition function:
$$Z_F(A) ={\displaystyle \int DV Z_B (A+V)e^{\tst -kS_{c.s.}(V)}}.$$

The bosonization formula is written in the second-quantized form
for the grand-canonical partition function:
$$\Xi = {\displaystyle \sum_N { e^{\beta \mu N} \over {N!}} Z_N},$$
and
\begin{equation}
\Xi_F (A) = {\displaystyle \int D\Psi\;D\Psi^* e^{\tst - S(\Psi, \Psi^*, A)}}
= {\displaystyle \int D\Phi\;D\Phi^*\; DV e^{\tst -[S(\Phi, \Phi^*, A+V)
+kS_{c.s.}(V)]}},
\label{bf}\end{equation}
where $\Psi$ is the Grassmann variable representing the fermion field,
while $\Phi$  is the complex variable representing the boson field.

There might be different choices of $G,\;k$:
In particular,
$G=U(1),\; k=1$ corresponds  to the abelian C-S bosonization\cite{sem},
while $G=U(1) \times U(1), \; k=(\theta, { {\theta +2} \over {\theta +1}})$
corresponds to the ``anyon'' bosonization\cite{ler}, with $\theta$ as
the statistics parameter. In this paper we will concentrate on the
case  $G= U(1) \times SU(2)$ with  $k = 2,1$, respectively\cite{fro,fm92}.

The relations for the correlation functions are given by

\begin{equation}
\Psi(x) \rightarrow P( e^{\tst i\int_{\gamma_x} V})\Phi (x),\;\;
\Psi^*(x) \rightarrow \; \Phi^* (x) P( e^{\tst - i\int_{\gamma_x} V})
\label{najw}\end{equation}
with $\gamma_x$  as a straight line  in the fixed time plane
joining $x$ with $\infty$.  This is a non-abelian generalization
of the  2D formula for the abelian case (\ref{2djw}).
It is important to notice the non-local character of these relations.

Here we have used the first-quantized
form of the path integral to identify the relation between
the fermion and boson systems, while the ``working'' formula for
bosonization is given in the second-quantized form of the path integral
representation. This switching from first to second quantized form
and vice versa will be frequently used throughout this paper.

\subsection{Outline of the Paper Content}
\label{cont}

The $U(1) \times SU(2)$ C-S bosonization approach has been successfully
employed by us earlier \cite{marc}
  to calculate the critical
exponents of the correlation functions in the 1D $t-J$ model in the limit
$t\gg J$.
Although, in principle,  all  bosonization schemes  should yield an exact
identity between
the correlation functions of the original fermionic field and corresponding
bosonic correlation functions, the  MFA, as mentioned earlier, gives
different results in
different bosonization schemes. The $U(1)$ C-S bosonization
has been shown to correspond essentially to the slave--boson and
slave--fermion approaches (depending on the choice of the gauge fixing);
while the non--abelian $U(1) \times SU(2)$ C-S  bosonization
corresponds to the slave semion--approach ($\theta = {1\over 4} $)\cite{fro}.
We have shown\cite{marc} that the ``semion'' spin-charge separation
of spinon and holon is the correct one to reproduce the exact exponents
known from the Bethe ansatz solution and the Luttinger liquid-conformal
field theory calculations\cite{ogata}.

	We considered 1D fermion system on the background of a 2D $U(1)
\times SU(2)$
C-S gauge field.  The $U(1)$ field is related to the charge, while the $SU(2)$
field is related to the spin degrees of freedom.
 Performing the dimensional reduction and using the freedom
in gauge-fixing, we could analyse the original problem as an
optimization process for the partition function of holons in a
spinon background. We could find an upper bound for the partition
function and an exact way to saturate this bound, without any approximations.
Afterwards we used MFA to consider an ``averaged'' holon configuration
to compute the long-time, large distance behavior of the correlation
functions, reproducing the exact results. The important lesson we
learned there is that the statistical (semionic) properties of
the constituent particles (spinons and holons) due to gauge field
fluctuations are crucial. In the operator form, the original fermion operator
can be decomposed as:
\begin{equation}
\psi_x = h_xe^{\tst  i{ \pi \over 2} {\dsp \sum_{l>x}} h^*_l h_l}[e^{\tst
i {\pi \over 2} {\dsp \sum_{l<x}} b^*_lb_l}b_x]
\label{scsep}\end{equation}
where $h_x$ is  a fermion, while $b_x$ is a hard core boson operator. However,
only $h_x$ together with the attached ``holon string'' (the exponentiated
operator)
represents a ``physical'' charged, spinless holon, whereas  $b_x$ together with
the attached ``spinon string'' corresponds to the neutral, spin ${1 \over
2}$ spinon.
Both of them satisfy the ``semionic'' equal-time commutation relataions:
$$
 f (x^1) f(y^1) = e^{\pm {i\pi \over 2}}  f (y^1)  f
(x^1), \qquad x^1  \mma y^1,
$$
where $f$ represents fermion (boson) operator along
with the attached string. This means that the ``semionic''
spinons and holons are ``deconfined'',
and the spin and charge are fully separated. The dynamics of the
spinon filed $b_x$ is described
 by an $O(3)$ NL$\sigma$-model. Due
to the presence of the topological term in 1D\cite{fra,gnt}, the spinons
are massless, and they are ``deconfined'', i.e., the spinons themselves, not
their bound states (the ususal spin waves in higher dimensions) are the
constituent quasiparticles.
These  results encourage us to explore the 2D case which is
of much more physical importance. We should, however, carefully distinguish
which are the generic features of the C-S gauge field theory under
consideration,
and what is specific for the 1D case.

In this paper we employ the $U(1) \times SU(2)$ C-S bosonization scheme to
study the 2D $t-J$ model in the underdoped regime in the limit $t \gg J$.
We will try to follow  as much as possible the same procedure as in 1D.
 The $U(1)$ gauge field $B$ is again  related to the charge
 degree of freedom, while the $SU(2)$ gauge field
$V$ is related to the spin degree of freedom.  First we prove the
existence of an upper bound of the partition function for  holons
in a spinon background, and we find the  optimal, holon-dependent spinon
configuration
which saturates the upper bound in an average sense.
The optimization arguments suggest coexistence of a flux--phase \cite{am}
with an  $s+i d$-like RVB state\cite{kot}, where the expectation value for the
Affleck--Marston(AM) bond--variable of spinons is close to 1, while
the $s+i d$-RVB order parameter is much smaller than 1.
Then we make an approximation,  neglecting
 the feedback of holon fluctuations on the
$U(1)$  field $B$ and spinon fluctuations on the $SU(2)$  field
$V$. Hence the  holon
field is  a fermion and the spinon field a hard--core boson.
Within this approximation we show that the $B$ field  produces a
$\pi$-flux phase for the holons, converting them into Dirac--like fermions,
while the $V$ field, taking into account the feedback of holons
  produces a gap for the spinons, minimal  at  $(\pm {\pi\over 2}, \pm
{\pi\over 2})$.
The spinons are described by a NL$\sigma$-model with a mass term (gap)
 $\simeq \sqrt{ -\delta \ln \delta}$, with $\delta$ as the doping
concentration.
This corresponds to a short-ranged  AF order (or disordered state in the
jargons of NL$\sigma$-model) in doped samples,  which  crosses over
to the long-ranged AF order in the prestine samples, when the gap vanishes.
To our knowledge,
this is the first successful attempt to include AF fluctuations
self-consistently in the RVB-type
approach. Moreover, we derive
a low--energy effective action in terms of spinons, holons and a
self-generated $U(1)$ gauge field. Neglecting  the gauge fluctuations,
the holons are described by  the   FL theory with a FS consisting of  4
``half-pockets'' centered at  $(\pm {\pi\over 2}, \pm {\pi \over
2})$ and  one  reproduces the results for the electron
spectral function obtained  in MFA (for the co-existing $\pi$-flux
 and $d$-wave RVB state)
\cite{dai}, in
qualitative agreement with the ARPES data \cite{mar}
 for underdoped cuprates.
If the gauge field were coupled to the spinons alone, it
would be confining, since the spinons are massive. However,
due to coupling to the  massless branch of holons (which are
actually non-relativistic because of a finite FS), gauge
fluctuations are \underbar{not} confining, but nevertheless yield an
attractive
interaction between spinons and holons leading to a bound state
in 2D with  electron quantum numbers.
This could   explain why neglecting the ``semionic'' nature of spinons and
holons is less  dangerous in 2D than in 1D. This means that the spin
and charge are not fully separated (like in 1D), showing up as bound states
in low-energy phenomena.
The renormalisation effects due to gauge
fluctuations  would lead to non--FL behaviour for the
composite  electron, in certain temperature range showing the linear
in $T$ resistivity discussed earlier\cite{ln,iw,marc1}.
This formalism provides a new interpretation of the spin gap in the
underdoped superconductors (mainly due to the short-ranged AF order)
and predicts that the minimal  gap for the physical electron
is proportional to the square root
of the doping concentration. Therefore
the gap does not vanish in any direction. All these predictions can be
checked explicitly in experiment.

The C-S gauge field approach has also been used  by Mavromatos and his
collaborators\cite{mav} to study the anyon superconductivity, advocated
by Laughlin\cite{lau}. To our understanding, the basic aim of their work
is to construct a model exhibiting semion superconductivity without
breaking the time-reversal and parity symmetry. In spite of some apparent
similarities in formulas, the main issue considered  and
the basic physical assumptions in their work are very
different from ours. They have also discussed the normal state properties
\cite{am1}, but the mechanism for a possible Non-FL behaviour in their paper
differs from what we consider here. We should also mention that the $SU(2)$
gauge field considered in their recent papers (quoted in
\cite{mav}) corresponds to a generalization of
the local $SU(2)$ symmetry at half-filling, and is not related to the spin
rotational symmetry, as we discuss in this paper. P.A. Lee and his
collaborators\cite{wen}
have also been considering  this (rather than the spin) $SU(2)$ group.

The present paper is an extended version of the earlier short
communication\cite{short}.

The rest of the paper is organized as follows:

In Sec. II we summarize the $U(1) \times SU(2)$ bosonization
in the context of 2D $t-J$ model;

In Sec. III we present the optimization problem for the spinon configuration;

In Sec. IV we derive the spinon effective action;

In Sec. V we consider the holon effective action;

In Sec. VI we make some concluding remarks.

The proof of the bound employed in Sec. III, is deferred to the Appendix.

\section{ $U(1) \times SU(2)$ Chern--Simons bosonization of the
$\lowercase{t}-J$ model}
\label{C-S-t-J}

\subsection{The Model Hamiltonian}
\label{model}

It is widely believed that the 2D $t-J$ model
captures the essential physical properties of the $Cu-O$ planes
characterizing a large class of high--$T_c$ superconductors\cite{anbook}.
The hamiltonian of the model is given by

\begin{equation}
H= P_G \Bigl[\sum_{<ij>} - t(\psi^\dagger_{i\alpha} \psi_{j\alpha} + h.c.)+
J \psi^\dagger_{i\alpha}
{\vec \sigma_{\alpha\beta} \over 2} \psi_{i\beta} \cdot
\psi^\dagger_{j\gamma} {\vec
\sigma_{\gamma\delta} \over 2} \psi_{j\delta} \Bigr] P_G,
\label{tj}
\end{equation}
where ${\vec \sigma_{\alpha\beta}}$ are the Pauli matrices,
 $\psi_{i\alpha}$ is the annihilation operator of a spin ${1\over 2}$
 electron on site $i$ of a square lattice, corresponding
to creating a hole on the $Cu$ site, and $P_G$ is the Gutzwiller projection
eliminating double occupation, modelling the strong on--site Coulomb
repulsion. To simplify notations, we introduce a two-component spinor
$\psi_i =( \psi_{i\uparrow}, \psi_{i\downarrow})$. Throughout this paper,
the small letters will denote operators, while the capital letters
will denote the corresponding complex ( or Grassmann) variables in the
path integral representation, unless otherwise specified.

Using the Hubbard-Stratonovich transformation
to introduce a complex gauge field $X_{<ij>}$,
 the grand--canonical partition function of the $t-J$ model at
temperature $T=1 / \beta$
and chemical potential $\mu$ can be
rewritten  as\cite{am}:

\be
\Xi_{t-J} (\beta, \mu) = \int {\cal D} X {\cal D}
X^* {\cal D} \Psi {\cal D} \Psi^* e^{\tst -S_{t-J} (\Psi, \Psi^*, X,
X^*)}
\label{tjgrand}
\ee
with

\bea
S_{t-J} (\Psi, \Psi^*, X, X^*) & = &
\int^\beta_0 d x^0 \left\{ \sum_{<ij>} \left({2 \over J} X^*_{<ij>} X_{<ij>} +
[(-t + X_{<ij>}) \Psi^*_{i \alpha} \Psi_{j \alpha}
+ h.c.]\right) \right. \nonumber\\
& &\left. + \sum_i \Psi^*_{i\alpha} (\partial_0 + \mu)
\Psi_{i\alpha}
+ \sum_{i,j} u_{i,j} \Psi^*_{i\alpha} \Psi^*_{j\beta} \Psi_{j\beta}
\Psi_{i\alpha}\right\},
\label{tjact}
\eea
where the two--body potential is given by

\be
u_{i,j} = \left\{\ba{ll} + \infty  \quad & i=j \cr
-{J \over 4} & i,j \ {\rm n.n.} \cr
0 & {\rm otherwise}.
\ea \right.
\label{uu}
\ee

Hereafter we denote the euclidean--time $x^0 \equiv \tau $
( $\partial_0 \equiv \partial / \partial \tau$)
 and its dependence of the fields
 is not explicitly spelled out.

\subsection{C-S Bosonization}
\label{cstj}

Comparing (\ref{tjgrand}), (\ref{tjact}) with (\ref{bf}),
 we find that the C-S bosonization
procedure, briefly introduced in  Sec. \ref{C-S basic} can be applied
to rewrite the grand-canonical partition function (\ref{tjgrand}). However,
there is an important difference, namely, the consideration in Sec.
\ref{C-S basic} was for 2D continuum, where the probability for
two world-lines (Brownian paths) to intersect each other at a given time is
zero.
This is not true for the lattice case we consider now, where the probability
for two paths to cross each other at a given time is not vanishing. On the
other hand,  the model we consider contains  a single-occupancy
constraint, expressed in terms of the Gutzwiller projection operator
$P_G$, or the infinite on-site repulsion $u_{i,i}$, which excludes
the intersection of paths. Therefore,
we can still apply the C-S bosonization scheme,
leaving the C-S gauge fields in the continuum, while
considering the matter field  on a discrete lattice.
We will introduce an abelian $U(1)$ gauge field $B$ related to
the charge degree of freedom and a $SU(2)$ gauge field $V$
related to the spin degrees of freedom.  The euclidean C-S actions for these
fields are given by:

\bea
& S_{c.s.} (B)= {1\over 4\pi i} \int d^3 x \epsilon^{\mu\nu\rho} B_\mu
\partial_\nu B_\rho, \nonumber \\
& S_{c.s.} (V)=  {1\over 4\pi i} \int d^3 x Tr \Bigl[ \epsilon^{\mu\nu\rho}
(V_\mu
\partial_\nu V_\rho + {2\over 3} V_\mu V_\nu V_\rho) \Bigr],
\label{cs}
\eea
where $V_\mu = V_\mu^a \sigma_a/2,\; a = 1, 2, 3, \;
\mu = 0, 1, 2 $ with  $\sigma_a$ as Pauli matrices.

In the fermion-boson transformation formula (\ref{bf}) $k=2$ for the $U(1)$
field $B$, and $k=1$ for the $SU(2)$ field $V$\cite{fro,fm92}. The correlation
functions for the Grassmann fields $\Psi_\alpha$ ($\Psi^*_\alpha$)
are substituted by the correlation functions of the gauge-invariant
complex fields $\Phi_\alpha(\gamma_x)$ ($\Phi^*_\alpha(\gamma_x$)),
defined as:

$$
\Phi_\alpha (\gamma_x) = e^{i\int_{\gamma_x} B} (P e^{i\int_{\gamma_x}
V})_{\alpha\beta} \Phi_{x\beta},
$$
$$
\Phi^*_\alpha (\gamma_x) = \Phi^*_{x\beta} (P e^{-i\int_{\gamma_x}
V})_{\beta\alpha} e^{-i\int_{\gamma_x} B}.
$$

As mentioned earlier, $\gamma_x$ is a straight line in the fixed-time
plane joining point $x$ with $\infty$ (reaching a compensating current
at $\infty$\cite{fro,fm92}) and $P$ is the path-ordering operator.
In principle, we can choose other gauge groups $G$ to implement
the bosonization scheme, but the encouraging result for the 1D $t-J$ model,
 reproducing the known exact exponents of the correlation functions\cite{marc}
 strongly favours the  $U(1)\times SU(2)$ choice.

The bosonized action is obtained via substituting the time derivative by the
covariant time derivative and the spatial lattice derivative by the
covariant spatial lattice derivative  in the $U(1) \times SU(2)$
bosonization

$$
\Psi^*_{j\alpha} \partial_0 \Psi_{j\alpha} \longrightarrow
\Phi^*_{j\alpha} \Bigl[\Bigl(\partial_0  + i B_0 (j)
\Bigr) \unity + i  V_0 (j) \Bigr]_{\alpha\beta} \Phi_{j\beta},
$$
$$
\Psi^*_{i\alpha} \Psi_{j\alpha} \longrightarrow \Phi^*_{i\alpha}
e^{i \int_{<ij>} B} (P e^{i \int_{<ij>} V})_{\alpha\beta}
\Phi_{j\beta}.
$$

We now decompose the bosonic field as follows:

\be
\Phi_{x\alpha} = \tilde E_x \Sigma_{x\alpha}, \ \Phi^*_{x\alpha}
= \tilde E^*_x \Sigma^*_{x\alpha},
\label{decomp}
\ee
where $\unity$ is a unity matrix, $\tilde E$  a complex scalar
field and $\Sigma_\alpha$ a complex
two--component spin ${1\over 2}$ field satisfying the constraint

\be
\Sigma^*_{x\alpha} \Sigma_{x\alpha} = 1.
\label{sigmaconst}
\ee

The gauge ambiguity involved in the decomposition (\ref{decomp}) will be
discussed later.  The field $\tilde E$ is coupled to the $U(1)$ gauge field
$B$ and
it describes the charge degrees  of freedom and the field $\Sigma_\alpha$
is coupled to the $SU(2)$ gauge field $V$ and it describes the spin degrees
of freedom of the original fermion. In this description the nature of
the groups associated with charge $\Bigl(U(1)\Bigr)$ and  spin
$\Bigl(SU(2)\Bigr)$
is explicitly exhibited and the coefficients of the C-S actions are
such that the charged and spin ${1\over 2}$ field operators reconstructed
from the gauge invariant (euclidean)
fields $(\tilde E_x e^{i\int_{\gamma_x} B}$ and
$P(e^{i\int_{\gamma_x} V})_{\alpha\beta} \Sigma_{x\beta})$ obey semionic
statistics\cite{fro,marc}.

In terms of $\tilde E$ and $\Sigma$ the $U(1) \times SU(2)$--bosonized action
of the $t-J$ model is given by

\begin{eqnarray}
& S_{t-J} (\tilde E, \tilde E^*, \Sigma, \Sigma^* , X, X^*, B,V) &\cr
& =\dsp \int_0^\beta dx^0 \Bigl\{   \dsp \sum_{<ij>} \Bigl( {2\over J}
X^*_{<ij>} X_{<ij>}
+  [(-t + X_{<ij>}) \tilde E_j^* e^{i \int_{<ij>}B} \tilde E_i
\Sigma^*_{i\alpha} (P e^{i \int_{<ij>} V})_{\alpha\beta} \Sigma_{j \beta}
 + h.c.] \Bigr) &\cr
&
+ \dsp \sum_j \left[ \tilde E^*_j (\partial_0 +
i B_0 (j) + \mu +{J \over 2}) \tilde E_j + \tilde E^*_j \tilde E_j
\Sigma^*_{j\alpha}
\Bigl(\partial_0  \unity + i
V_0 (j)
\Bigr)_{\alpha\beta} \Sigma_{j\beta}\right] &\cr
&+ \dsp \sum_{i,j} u_{i,j} \tilde
E_i^* \tilde E_i \tilde E_j^* \tilde E_j\Bigr\} + 2
S_{c.s.} (B) + S_{c.s.} (V) &
\label{bosonac}
\end{eqnarray}
with constraint (\ref{sigmaconst})  and Coulomb gauge-fixing
for the $U(1) \times SU(2)$ field implemented \cite{marc}.

It is convenient to describe the charge properties in terms of a hole--like
field $H$ (holon) and this can be achieved in this formalism by
substitution $\tilde E \rightarrow H^*, \tilde E^* \rightarrow H$, with $H,
H^*$ as Grassmann fields, and changing the sign of the C-S action
for the $B$ field \cite{fm92,marc}. After integration over the
auxiliary gauge field $X$, the
grand--canonical partition function $\Xi (\beta, \mu)$ can be rewritten
as:

\be
\Xi (\beta,\mu) = \int {\cal D} H {\cal D} H^* {\cal D} \Sigma_\alpha
{\cal D} \Sigma_\alpha^*
{\cal D} B {\cal D} V e^{\tst -S (H, H^*, \Sigma, \Sigma^*, B, V)} \delta
(\Sigma^* \Sigma -1),
\label{fermigran}
\ee
where the euclidean action is given by:

\begin{eqnarray}
&S (H, H^*, \Sigma, \Sigma^*, B, V)
=\dsp \int^\beta_0 dx^0 \Bigl\{ \dsp\sum_j \Bigl[ H^*_j \Bigl( \partial_0 -
i B_0 (j)
- (\mu + {J\over 2})\Bigr) H_j
+i B_0 (j) &\cr
& + (1 - H^*_j H_j) \Sigma^*_{j\alpha} \Bigl(\partial_0 + i V_0 (j)
\Bigr)_{\alpha\beta} \Sigma_{j\beta}\Bigr]
+ \dsp\sum_{<ij>}\Bigl[ (- t H^*_j e^{i\int_{<ij>} B} H_i
\Sigma^*_{i\alpha} (P
e^{i\int_{<ij>}V})_{\alpha\beta} \Sigma_{j\beta}  + h.c.) & \cr
&   + {J\over 2} (1 - H^*_j H_j) (1 - H^*_i H_i) \Bigl(|\Sigma^*_{i\alpha}
(P e^{i \int_{<ij>} V})_{\alpha\beta}
\Sigma_{j\beta} |^2 - {1\over 2} \Bigr) \Bigr] \Bigr\}
- 2 S_{c.s.} (B) + S_{c.s.} (V) .
\label{fermiac}
\end{eqnarray}
 In what follows we will denote the shifted chemical potential
$ \mu' = \mu + J/2$ by $\delta$, proportional to the doping
concentration.

\subsection{Gauge Fixings}
\label{gf}

The action (\ref{fermiac}) is invariant under the local  gauge transformations:

\bea
 U (1) : & H_j \rightarrow H_j e^{i\Lambda (j)},\;\; H^*_j \rightarrow H^*_j
E^{-i\Lambda (j)},& \nonumber\\
& B_\mu (x) \rightarrow B_\mu (x) + \partial_\mu \Lambda
(x), \hspace*{2truecm} &\Lambda (x) \in {\bf R} \nonumber\\
 SU (2) :& \Sigma_j \rightarrow R^\dagger (j) \Sigma_j,\;\; \Sigma_j^*
\rightarrow \Sigma^*_j R(j),&\nonumber \\
& V_\mu (x) \rightarrow R^\dagger (x) V_\mu (x) R
(x) + R^\dagger (x) \partial_\mu R(x),& \;\; R(x) \in
SU(2)
\label{gauge}
\eea
and an additional holon-spinon $(h/s)$ gauge invariance arising from the
ambiguity in the
decomposition (\ref{decomp}):

\bea
h/s : & H_j \rightarrow H_j e^{i\zeta_j},\;\; H_j^* \rightarrow H^*_j
e^{-i\zeta_j}, & \nonumber \\
& \Sigma_{j\alpha} \rightarrow \Sigma_{j\alpha} e^{i\zeta_j}, \;\;
\Sigma^*_{j\alpha} \rightarrow \Sigma^*_{j\alpha} e^{-i\zeta_j},&
\;\;\;\zeta_j \in
{\bf R}.
\label{hsgauge}
\eea

\vskip 0.3truecm

It is important to remark that
the theory in terms of $\{H, H^*, \Sigma, \Sigma^*, B, V\}$ is equivalent to
the original fermionic theory only if the $h/s$ gauge is fixed  to respect
$U(1) \times SU(2)$ invariance. The $h/s$ gauge--fixing will be discussed
later.

\smallskip
We first gauge--fix the $U(1)$ symmetry imposing a Coulomb condition on $B$
(from now on $\mu =1,2)$:

\be
\partial^\mu B_\mu =0.
\label{u1cou}
\ee

To retain the bipartite lattice structure induced by  the AF interactions,
 we gauge--fix  the $SU(2)$ symmetry  by a ``N\'eel gauge'' condition:

\be
\Sigma_j = \sigma_x^{|j|} {1\choose 0}, \;\;\; \Sigma^*_j = (1,0)
\sigma_x^{|j|},
\label{neel}
\ee
where $|j|= j_1 + j_2$.  Then we split the integration over $V$ into an
integration over a field $V^{(c)}$, satisfying the Coulomb condition:

\be
\partial^\mu V_\mu^{(c)} =0,
\label{su2cou}
\ee
and its gauge transformations expressed in terms of an $SU(2)$--valued
scalar field $g$ (not a second-quantized operator), i.e.,
$V_a =g^\dagger V_a^{(c)} g + g^\dagger\partial_a g, \; a = 0,1,2.$

Integrating over $B_0$, we obtain

\be
B_\mu = \bar B_\mu + \delta B_\mu,\;\;\; \delta B_\mu (x) = {1\over 2}
\dsp \sum_j H^*_j H_j \partial_\mu {\rm arg} \ (x-j),
\label{u1fix}
\ee
where $\bar B_\mu$ gives rise to a $\pi$-flux phase, i.e.,
$e^{i\int_{\partial p} \bar B} = -1$ for every plaquette $p$.

Integrating over $V_0$, we find

\be
V_\mu^{(c)} = \dsp\sum_j (1- H^*_j H_j) (\sigma_x^{|j|} g^\dagger_j
{\sigma_a \over 2} g_j \sigma^{|j|}_x)_{11} \partial_\mu {\rm arg} \
(x-j) \sigma_a,
\label{su2fix}
\ee
where $\sigma_a, a= x, y, z$ are the Pauli matrices.
After the $U(1)\times SU(2)$ field being gauge--fixed, as discussed above,
the action (\ref{fermiac}) becomes

\bea
&S(H, H^*, g) =\dsp\int^\beta_0 d x^0 \Bigl\{ \sum_j\Bigl[ H^*_j
\Bigl(\partial_0 -\delta
\Bigr) H_j
+ (1 - H^*_j H_j)  (\sigma_x^{|j|} g^\dagger_j \partial_0 g_j
\sigma_x^{|j|})_{11}\Bigr] &\cr
& +\dsp \sum_{<ij>}  \Bigl[- t H^*_j e^{i \int_{<ij>} \bar B +\delta B} H_i
(\sigma_x^{|i|} g^\dagger_i (P e^{i \int_{<ij>} V^{(c)}}) g_j
\sigma_x^{|j|})_{11} + h.c.\Bigr] &  \cr
&+ {J\over 2}\dsp\sum_{<ij>} (1 - H^*_i H_i) (1 - H^*_j H_j) \Bigl[|
(\sigma_x^{|i|}
g^\dagger_i  (P e^{i\int_{<ij>} V^{(c)}}) g_j \sigma_x^{|j|})_{11}|^2 -
{1\over 2} \Bigr]\Bigr\}. &
\label{fermiac1}
\eea

Here the boundary terms are omitted and
the $ S^1 \times {\bf R}^2$ topology
of the involved euclidean space--time imposes the vanishing of the
topological term $Tr \int_{D\times {\bf R}^2} (g^\dagger dg)^3$, where $D$ is
a disk of radius $\beta$.

Equation (\ref{fermiac1}) is the starting point for our subsequent
analysis.
In $S(H, H^*, g)$ the charge degrees of freedom (d.o.f.) are described by $H,
H^*$ (2 d.o.f.) and the spin degrees of freedom by $g$ (3 d.o.f.)
subjected to a
constraint (-1 d.o.f.) coming from the $h/s$ gauge fixing,
reproducing the correct
counting of degrees of freedom of the original fermionic fields
$\Psi_\alpha, \Psi^*_\alpha$ (2+2 d.o.f.) in the euclidean path--integral
formalism.


\section{ optimization of the spinon configuration}
\label{optim}
 To analyse (\ref{fermiac1}) we first recall the strategy  adopted
for the 1D case\cite{marc}. We noticed that
one can find an upper  bound for the  partition function
of holons in a spinon background. Moreover,
one can find explicitly the spinon configuration, exactly saturating this
bound.
Then one can consider the quantum fluctuations around this optimal
configuration to evaluate related physical quantities. Here we will follow
a similar strategy, namely to search first for the upper bound of the
partition
function for holons in a spinon background. It turns out that such
an upper bound exists. However,
unlike the 1D case, we cannot find a  spinon configuration exactly saturating
this upper bound. Nevertheless, we can find a holon-dependent spinon
configuration
$g^m$ which is optimal, saturating the upper bound on average,
 and take it as the starting point to consider
 the spinon fluctuations.

\subsection{Auxiliary Lattice Gauge Field and Upper Bound}
\label{auxi}

To find the optimal configuration we introduce an
auxiliary lattice gauge field $\{A, U \}$,
with $A_j \in {\bf R}$ (real),  $U_{<ij>} \in {\bf C}$ (complex),
$|U_{<ij>}|\leq 1$, and an action $S= S_1 + S_2,$

\bea
&S_1 (H, H^*, A, U) & \cr
& = \dsp \int^\beta_0 d x^0\Bigl\{\sum_j\Bigl[ H^*_j \Bigl(\partial_0 -
\delta \Bigr) H_j
+ i (1 - H^*_j H_j) A_j\Bigr] + \sum_{<ij>} (-t H^*_i U_{<ij>} H_j +
h.c.)\Bigr\},&\cr
&S_2 (H, H^*, U) = \dsp \int^\beta_0 dx^0 \sum_{<ij>} {J\over 2} (1 - H^*_i
H_i)
(1 - H^*_j H_j)
\Bigl(|U_{<ij>}|^2 - {1\over 2} \Bigr).&
\label{auxiac}
\eea

The action (\ref{auxiac}) equals (\ref{fermiac1}) if we make the
identifications:

\bea
& & i A_j \sim (\sigma_x^{|j|} g^\dagger_j \partial_0 g_j
\sigma_x^{|j|})_{11}, \cr
& & U_{<ij>} \sim  e^{\tst -i \int_{<ij>}( \bar B + \delta B) }
(\sigma_x^{|i|}
g^\dagger_i (P e^{\tst i\int_{<ij>} V^{(c)}} )g_j \sigma_x^{|j|})_{11},
\label{iden}
\eea
(but in the derivation of the following bound these identifications are not
made.)

Let

$$
\Xi (A, U) = \int {\cal D} H {\cal D} H^* e^{\tst -S(H, H^*, A,U)},
$$
  we prove in Appendix A the following upper bound:

\be
|\Xi (A, U) | \leq  \int {\cal D} H {\cal D} H^* e^{\tst -[S_1 (H,H^*, 0, \hat
U)+ S_2 (H, H^*, 0)]},
\label{bound}
\ee
where $\hat U$ is the time-independent $U$-configuration maximizing

\be
\int {\cal D} H {\cal D} H^* e^{\tst -[S_1 (H, H^*, 0, U) + S_2 (H, H^*,
0)]}|_{\partial_0 U=0}.
\label{bound1}
\ee

To discuss the properties of $\hat U$ we first notice that the quantity
optimized by $\hat U$ is the free energy $F(U)$ of a gas of spinless holes
at temperature $T= \beta^{-1}$ with chemical potential $\delta\;
 (\;\delta=0$ corresponding to half--filling) on a lattice, with hopping
parameter on  the link $<ij>$ given by $t|U_{<ij>}|$, in the presence of a
constant (but not uniform) magnetic field where the flux through a plaquette
$p$ is given by arg $(U_{\partial p})$ and is subjected to an attractive n.n.
density--density interaction with coupling constant ${ J\over 4}$. We
consider the
system at large $\beta$, and small $\delta$ and make the following:

\smallskip
\underbar{Assumptions:} 1) we consider negligible the density--density
interaction, since the holon density $(\delta)$ is small;

\qquad\qquad\quad\quad  2) we assume translational invariance of $\hat U$.

\vskip 0.3truecm
\underbar{Remark:} Assumption 2) appears to be reasonable
in the light of the results
of \cite{ll}, where  it has been shown that, the
configuration $U$ maximizing the determinant of the hopping matrix of the
above system is translation invariant.

By gauge--invariance (see Appendix A) the result of optimization depends
only on $|\hat U_{<ij>}|$ and arg $(\hat U_{\partial p})$.
As a consequence of the assumption 2), $F(U)$ is monotonically
increasing in $|U|$, hence

\be
|\hat U_{<ij>}| =1.
\label{unimod}
\ee

It has been conjectured in \cite{has} and proven in \cite{br} that the
ground state
energy at $T=0$
of the system under consideration is optimized in the magnetic field
chosen as:

\be
{\rm arg} \ (\hat U_{\partial p}) = \pi (1 - \delta)
\label{phase}
\ee
which is the commensurability condition for the flux.
It is then natural to conjecture that this remains true for  large
enough $\beta$.

\subsection{Optimal Spinon Configuration}
\label{saturation}

In the last subsection we have stated a kind of ``theorem'' (proven in
Appendix A),
now we will find out the consequences of this theorem in our context,
namely, assuming (\ref{unimod}) and  (\ref{phase}) to hold, we  attempt to find
a holon--dependent spinon configuration $(g^m)$
saturating the bound (\ref{bound}), using the
identifications (\ref{iden}).

Following a strategy developed in 1D \cite{marc}, we introduce a
first--quantized
(Feynman-- Kac) representation of the holon partition function in the
presence of a $g$--background. As in 1D, a key ingredient in the analysis is
that, for  a fixed link $<ij>$, in
every holon configuration the term $(\sigma_x^{|i|} g^\dagger_i P
e^{i\int_{<ij>} V^{(c)}} g_j \sigma_x^{|j|})_{11}$ appears, in the first
quantized formalism, either in the worldlines of
holons or in the Heisenberg term, but never simultaneously. This is
the consequence of the single-occupancy constraint and permits a
separate optimization of $S_1$ and $S_2$, as required in the bound
(\ref{bound}). It
turns out, however, that, contrary to the 1D results, we cannot identify a
specific configuration $g^m$ saturating the bound exactly, but only
approximatively, in an appropriate average sense.

The whole procedure in 2D can then be justified in the limit $t >> J$
because the effective mass of holes is very heavy, as a result of the large
number of soft spinon fluctuations surrounding the hole in its
motion\cite{ly}
and in a sense our treatment can be considered as a kind of Born--Oppenheimer
approximation for the spinons in the presence of the holons.

Here we do not give  details of the derivation, but rather introduce
notations and quote the obtained results. Those interested in further
discussions
are referred to \cite{gin,fro,fm92,marc}.

Let $\Delta$ denote the 2D lattice laplacian defined on a scalar lattice
field $f$ by

$$
(\Delta f)_i = \sum_{j:|j-i|=1} f_j - 4 f_i;
$$
let $d\mu (\omega)$ denote the measure on the random walks $\omega$ on the
2D lattice such that

$$
\Bigl(e^{\beta\Delta} \Bigr)_{ij} = \int_{\scriptstyle \omega(0)= i\atop
\scriptstyle\omega(\beta)=j} d\mu (\omega),\;\; \beta >0;
$$
let $P_N$ be the group of permutations of $N$ elements and, for $\pi \in P_N$,
let $\sigma (\pi)$ denote the number of exchanges in $\pi$, then the
partition functions of holons $(H)$ in a given $g$ background can be
rewritten as:

\bea
&\Xi (g) = e^{\tst i\sum_j \int^\beta_0 dx^0 A_j}\dsp \sum^\infty_{N=0}
{e^{\beta\delta N} \over N!}\dsp \sum_{\pi \in P_N} (-1)^{\sigma(\pi)}
\dsp\sum_{j_1 ... j_n}  \dsp \prod^N_{r=1} \int_{\scriptstyle\omega_r (0) =
j_r\atop\scriptstyle\omega_r (\beta)= j_{\pi (r)}} d\mu
(\omega_r) & \cr
& \cdot \dsp \prod_{<ij> \in{\underline{\omega}^\bot}} t U_{<ij>}
 e^{-\tst i\int_{\underline{\omega}^\Vert} dx^0 A} e^{\tst -\sum_{<ij> \cap
\underline{\omega} = \emptyset}
{J\over 2}\tst \int^\beta_0 dx^0 (|U_{<ij>}|^2 - {1\over 2} )}, &
\label{gpart}
\eea
where identifications (\ref{iden}) are understood and, for a fixed $N$,
${\underline
{\omega}} = \{\omega_1, ... ,\omega_N\}$ denote the worldlines of holon
particles, ${\underline{\omega}^\bot}$ the components of
${\underline{\omega}}$
perpendicular and ${\underline{\omega}^\Vert}$ parallel to the time axis,
respectively.

To saturate the bound (\ref{bound}) with a configuration $g^m
({\underline{\omega}})$ we first impose

\be
i A_j = (\sigma_x^{|j|} g^\dagger_j \partial_0 g_j \sigma_x^{|j|})_{11} =0,
\quad j \in {\underline{\omega}^\parallel}.
\label{paral}
\ee
 Eq. (\ref{paral}) is satisfied choosing $g^m$ constant during the period
when no particle hops.

Imposing  $U_{<ij>} =0$ in $S_2$ (see (\ref{auxiac})) corresponds in the first
quantized formalism to setting:

\be
(\sigma^{|i|}_x g^\dagger_i P e^{i\int_{<ij>} V^{(c)} } g_j
\sigma_x^{|j|})_{11} = 0, \;\;\; <ij> \cap \, \underline{\omega} = \emptyset.
\label{perp}
\ee
In physical terms this means that the $s+ id$ RVB order parameter\cite{kot}
is very small (see the discussion at the end of Sec. IV).

We notice that if

\be
g_j = \cos \theta_j \unity +  i \sin  \theta_j \sigma_z, \;\;\; j \notin
{\underline{\omega}}
\label{gperp}
\ee
for some angle $\theta_j \in [0, 2\pi)$, (\ref{perp}) is then satisfied. In
fact, since $V^{(c)}$ depends only on sites where there are no holes ( see
(\ref{su2fix})), from (\ref{gperp}) it follows that correspondingly

\be
V^{(c)} (x) = \sum_j (1 - H_j^* H_j) {(-1) \over 2}^{|j|} \partial_\mu
{\rm arg} \ (x-j) \sigma_z,
\label{vc}
\ee
so that $g^\dagger_i P e^{i\int_{<ij>} V^{(c)} } g_j$ has only diagonal
components.

This result shows that in the N\'{e}el gauge, quite independently of doping
concentration (since the condition $U=0$ in $S_2$ does not depend on small
doping assumption), one should expect that the physics is dominated by
$V^{(c)}$ only in the $U(1)$ subgroup of $SU(2)$ related to the axis
chosen in the N\'{e}el gauge.

The condition $|\hat U_{<ij>}|=1$  in $S_1$ (see (\ref{auxiac}))
corresponds to imposing:
\be
|(\sigma_x^{|i|} g^\dagger_i P e^{\tst i\int_{<ij>} V^{(c)}} g_j
\sigma_x^{|j|})_{11}|=1, \;\;\; <ij> \in {\underline{\omega}}^\bot
\label{uperp}
\ee
which means in physical terms that the AM order parameter\cite{am}
is of the order 1 (see the discussion at the end of sec. IV).

To discuss (\ref{uperp}) we recall that the paths on which $d\mu (\omega)$ is
defined are left--continuous \cite{marc}, so that
at the jumping time $\tau, \omega_r (\tau) = \lim_{\epsilon\searrow 0}
\omega_r (\tau +\epsilon)$,
or, in simpler terms, one should think the holon at $\tau$ at the end of
the jumping link, oriented according to the increasing worldline time of
the holon. As a consequence, if $<ij> \in {\underline{\omega}}^\bot$,
either $i\in {\underline{\omega}}$ or $j \in {\underline{\omega}}$, but
never both. Let us assume $j\in {\underline{\omega}}$, then according to
the previous requirements
$$
g^\dagger_i P e^{\tst i\int_{<ij>} V^{(c)} } = \cos  \theta_{<ij>} \unity
+ i \sin  \theta_{<ij>} \sigma_z,
$$
for some angle $\theta_{<ij>} \in [0, 2\pi)$. We represent
$$
g_j = \cos \varphi_j + i \vec\sigma \cdot \vec n_j \sin
\varphi_j,
$$
for some unit vector $\vec n_j$ and angle $\varphi_j \in [0, 2\pi)$.

From (\ref{uperp}) one immediately obtains

\be
\varphi_j = {\pi\over 2}, \quad n_{j z} =0.
\label{gi}
\ee

Finally, let us try to impose the condition (\ref{phase}).
This translation invariant condition cannot be exactly fulfilled (unlike
 in 1D where a configuration $g^m$ exactly saturating the
bound analogous to (\ref{bound}) can be found). However, we notice that the
$B$--dependent part of arg $(U_{\partial p})$ has a translation
invariant mean
satisfying the above condition, ( see (\ref{bound}) and (\ref{u1fix})),
hence it is natural
to impose (on average):

\be
\Bigl(\sigma_x^{|i|} g^\dagger_i e^{\tst i\int_{<ij>} V^{(c)}} g_j
\sigma_x^{|j|}\Bigr)_{11} \simeq 1, \;\;  <ij> \in {\underline{\omega}}^\bot.
\label{hop}
\ee

Defining

\be
\bar g_j = e^{\tst -{ i\over 2}\dsp \sum_{\ell \not = j} (-1)^\ell \sigma_z
{\rm arg} \
(\ell - j) },
\label{gbar}
\ee
and choosing

\be
g_j = \left\{ \begin{array}{ll}
\bar g_j, & \qquad j \notin {\underline{\omega}} \cr
\bar g_j \tilde g_j,  & \qquad j \in {\underline{\omega}}
\end{array}\right.
\label{gdef}
\ee
we can kill the fast fluctuating first term in (\ref{vc}). The remaining term,
denoted by
$$\bar V=- \sum_j  H_j^* H_j {(-1) \over 2}^{|j|} \partial_\mu
{\rm arg} \ (x-j) \sigma_z, $$
 for small hole concentration is a slowly varying field
yielding a contribution $O(\delta)$ to $\arg (U_{<ij>}),$ with zero
translational average. The final result is that we can assume for the
optimizing configuration, using (\ref{gi}), (\ref{gbar}), (\ref{gdef})

$$
g^\dagger_i e^{\tst i \int_{<ij>} V^{(c)}} g_j = e^{\tst i \int_{<ij>} \bar
V} \tilde g_j
\simeq \tilde g_j = e^{\tst i {\pi\over 2}(\sigma_x n_{jx} + \sigma_y n_{j
y})},
$$
and we immediately derive from (\ref{hop}) the condition $n_{j x} =0, n_{jy} =
(-1)^{|j|}$.

Notice that from these definitions:

\be
\tilde g_j \sigma_x^{|j|} = \sigma_x^{|j|+1}.
\label{gtilde}
\ee

In view of the optimization discussed  above it appears natural to
introduce a variable $R_j \in SU(2)$ describing spinon fluctuations around
the optimizing configuration, through the definition

\be
g_j = \bar g_j R_j \tilde g_j= e^{\tst -{ i\over 2}\dsp \sum_{\ell \not = j}
(-1)^\ell \sigma_z {\rm arg}  (\ell - j) }R_j e^{ \tst i {\pi \over 2}
(-1)^{|j|}\sigma_y H^*_jH_j}
\label{rdef}
\ee
with $R_j$ being represented in $CP^1$ form as

\be
R_j = \pmatrix{b_{j1} & b^*_{j2}\cr
b_{j2} & b^*_{j1} \cr} , \;\; b^*_{j\alpha} b_{j\alpha} =1.
\label{rmat}
\ee
where $b_{j\alpha}$ is a two--component complex field. The optimal
configuration $g^m$
is given by $R =\unity$.

Using eq. (\ref{gtilde}) and the $SU(2)$--gauge invariance of the (formal)
measure ${\cal D} g$ to
absorb $\bar g$, the partition function of the $t-J$ model can be
exactly rewritten in terms of the euclidean action $S= S_h + S_s$,

\bea
& S_h = \dsp  \int^\beta_0 dx^0 \Bigl\{ \sum_j H^*_j [\partial_0
-(\sigma_x^{|j|} R^\dagger_j
\partial_0 R_j \sigma_x^{|j|})_{11}
- \delta] H_j & \cr
& + \dsp \sum_{<ij>} [-t H^*_j e^{\tst - i \int_{<ij>} (\bar B +
\delta B) } H_i (\sigma_x^{|i|} R_i^\dagger P e^{\tst i\int_{<ij>} (\bar V
+\delta
V) } R_j \sigma_x^{|i|})_{11} + h.c. ]\Bigr\}, &
\label{sh}
\eea

\bea
& S_s = \dsp \int^\beta_0 dx^0 \Bigl\{ \sum_j (\sigma_x^{|j|} R^\dagger_j
\partial_0 R_j
\sigma_x^{|j|})_{11} & \cr
& +\dsp \sum_{<ij>} {J\over 2} (1 - H^*_i H_i) (1- H^*_j H_j)
\left[|(\sigma_x^{|i|}
R^\dagger_i P e^{\tst i\int_{<ij>} (\bar V + \delta V) } R_j
\sigma_x^{|j|})_{11}
|^2 - {1\over 2} \right]\Bigr\},&
\label{ss}
\eea
where $\delta V = V^{(c)} - \bar V$.

Let us remark again that in (\ref{sh}), (\ref{ss}) no approximations have
been made.
The point of the above analysis is that for large enough $\beta$ and
 small enough $\delta$ we expect that $R$ and $\delta V$ describe small
fluctuations.

\section{Spinon effective action}

\subsection{ The Main Approximation}

To proceed further we make the following

\underbar{Approximation}
$(\delta V = \delta B =0):$
we assume that the spinon fluctuations $(R)$ are
small enough (for $\beta$ large, $\delta$, $J/t$ small)
that we can neglect their
back reaction on the gauge field $V$, i.e. we set $\delta V=0$.
We expect that the main effect of the neglected fluctuations of $V$ is to
convert the gauge invariant spinon field operator reconstructed from the
euclidean field

$$
(P e^{\tst i\int_{\gamma_j} V} ) \Sigma_j = e^{\tst i\int_{\gamma_j} (\bar
V + \delta
V) } \bar g_j R_j \sigma_x^{|j|} {1 \choose 0}
$$
into a semion field operator. Retaining the fermionic nature of $\Psi$ is
then inconsistent with neglecting $\delta V$ (which also  introduces a
fictitious parity breaking) unless we neglect also $\delta B$, responsible
for the semionic nature of the gauge invariant holon field operator
reconstructed from the euclidean field

$$
e^{\tst -i \int_{\gamma _j} (\bar B +\delta B) } H_j.
$$

In 1D the proper account of the statistics of  the holon and spinon field
operators was crucial for deriving ( within the C-S approach) the
correct physical properties of the model, known by Luttinger
liquid and conformal field theory techniques \cite{ogata}. However,
 in 2D we believe the statistics
of holons and
spinons is less crucial because we expect that, contrary to 1D, they
form a bound state, as will be discussed later.

To derive the low--energy spinon action let us start computing the link
variable
\be
R^\dagger_i e^{\tst i\int_{<ij>} \bar V} R_j = \pmatrix{\alpha_{<ij>} b^*_{i1}
b_{j1} + \alpha^*_{<ij>} b^*_{i2} b_{j2} & -\alpha_{<ij>} b^*_{i1} b^*_{j2}
+ \alpha^*_{<ij>} b^*_{i2} b^*_{j1} \cr
-\alpha_{<ij>} b_{i2} b_{j1} + \alpha^*_{<ij>} b_{i1} b_{j2} &
\alpha_{<ij>} b_{i2} b^*_{j2} + \alpha^*_{<ij>} b_{i1} b^*_{j1} \cr},
\label{link}
\ee
where $\alpha_{<ij>} = e^{\tst {i\over 2} \int_{<ij>} \bar V_z}$. Looking back
at equation (\ref{ss}) we find that in the hopping term of holons only the
diagonal elements of (\ref{link}) appear, a kind of gauge--invariant
 AM variable\cite{am}, whereas in the Heisenberg term only
the off--diagonal elements of (\ref{link}) appear, a kind of gauge--invariant
RVB variable\cite{kot}.
According to the optimization arguments given in the previous Section, the
vacuum expectation value of the AM gauge variable is expected to be
$s$-like, real and close to 1 ( see eq. (\ref{hop})), while the RVB order
parameter should be rather small( see eq. (\ref{perp})). These anticipations
are fully confirmed by the mean field calculations\cite{dai,sheng}.

\subsection{Nonlinear $\sigma$ Model with Mass Term}

We now derive a low--energy continuum effective action for spinons by
rescaling the model to a lattice spacing $\epsilon <<1$ and neglecting
higher order terms in $\epsilon$. As it is standard in AF systems\cite{fra},
 we  assume
\be
b^*_{j\alpha} \vec\sigma_{\alpha \rho} b_{j\beta} \sim \vec\Omega_j +
(-1)^{|j|} \epsilon \vec L_j,
\label{stag}
\ee
with $\vec\Omega^2_j = f \lqua 1,\;  \vec\Omega \cdot \vec L =0,$ where
$\vec\Omega, \vec L$ are defined on a sublattice e.g. $\vec\Omega_j
\equiv \vec\Omega_{j_1+{1\over 2}, j_2}$, $\vec L_j \equiv \vec L_{j_1 +
{1\over 2}, j_2}, j_1 = j_2$ mod 2 and they describe the AF and
ferromagnetic fluctuations, respectively.
It is  useful to rewrite $\vec\Omega$ in the $CP^1$ form:

\be
\vec\Omega = z^*_j \vec\sigma_{\alpha\beta} z_\beta, \qquad
z^*_\alpha z_\alpha = f,
\label{cp1}
\ee
with $z_\alpha, \alpha = 1,2$ as a spin ${1\over 2}$ complex (hard-care)
boson field. Consistently with the slowly--varying nature of $\bar V$ for
small hole concentration, we assume

\be
e^{\tst -i\int_{<j\ell>} \bar V_z} -1 \sim \epsilon (-i \bar V_z) (j) +
{\epsilon^2\over 2} (-i \bar V_z)^2 (j) + O (\epsilon^3).
\label{expa}
\ee

On the rescaled lattice the Heisenberg term becomes

\bea
&\dsp {J\over 2}  \sum_{<ij>}  {|(\sigma_x^{|x|} R^\dagger_i e^{\tst
{i\over 2}
\int_{<ij>} \bar V} R_j \sigma_x^{|j|})_{11}|^2 - {1\over 2} \over
\epsilon^2 }  & \cr
& = \dsp {J\over 2} \sum_{<ij>}  \left\{ {1\over 2} \Bigl({\vec\Omega_i -
\vec \Omega_j
\over \epsilon}\Bigr)^2 + 2\vec L^2_j + \bar V^2_z (j) \Bigl[(\Omega_{jx})^2 +
(\Omega_{jy})^2 \Bigr]\right\} + O (\epsilon).&
\label{expan}
\eea

For the temporal term  we obtain analogously

\be
- (\sigma_x^{|j|} R^\dagger_j \partial_0 R_j \sigma_x^{|j|})_{11} = \sum_j
(-1)^{|j|} z^*_{j\alpha} \partial_0 z_{j\alpha} +
{\epsilon\over 2} \vec L_j \cdot (\vec\Omega_j \wedge \partial_0
\vec\Omega_j) +
O (\epsilon^2).
\label{texp}
\ee

The first term in (\ref{texp}) actually vanishes, since otherwise
it would produce a topological
$\theta$--term which is known to be absent in 2D\cite{fra}.

In treating the holon density terms in the spinon action in MFA,
 we keep only the leading fluctuation terms and neglect terms
of the order  of $O (J\delta), O (\delta^2)$. Then,
integrating out $\vec L$ and taking the continuum limit, from (\ref{stag}) --
(\ref{texp}) we obtain the NL$\sigma$-model  effective
action for spinons $S^\star_s + S'_s$,

\be
S^\star_s = \int d^3 x {1\over g} [(\partial_0 \vec\Omega)^2 + v^2_s
(\partial_\mu \vec\Omega)^2 +(\vec\Omega)^2 (\bar V_z)^2], \quad S'_s =
- {1\over g} \int d^3 x \Omega_z^2 \bar V^2_z,
\label{spinac}
\ee
where the coupling constant $g$ and the spin wave velocity
$v_s$ are easily derived functions of $J,\, \delta,\, \epsilon$.

We now make the \underbar{Approximation P:} we treat the term $S'_s$ as a
perturbation. To understand the physics described by $S^\star_s$ we first
notice that if $\bar V_z^2$ were absent, the NL$\sigma$-model would be in
the symmetry broken  phase, since $g$ is small $(\sim J)$ at the lattice
scale. For larger scales $g_{eff}$ then flows towards its critical value,
describing the large--distance properties of a NL$\sigma$ model with an
insulating ``N\'{e}el'' ground state and spin--wave Goldstone
excitations\cite{fra}. To
get an idea of the effect of $\bar V^2_z$ we replace the NL$\sigma$ model
constraint $\vec\Omega^2 =f$ by a softened version, adding to the
lagrangian a term $\lambda (\vec\Omega^2 - f)^2$, expected to produce the
same low--energy behaviour, and we replace $\bar V^2_z$, a function of the
holon positions, by its statistical average $<\bar V^2_z>$. By its
definition,  $\bar
V^2_z$ is positive definite and we give a rough estimate of $<\bar V^2_z>$ by
first performing a translational average at a fixed time over a fixed holon
configuration and then an average over holon configurations with mean holon
density $\delta$.

In the first step let ${\underline{x}} =\{x_i\}^N_{i=1}$ denote the holon
positions and let $q = \{q_i\}^N_{i=1}, q_i =(-1)^{|x_i|+1}$ and
restrict the computation to a finite volme $|\Lambda|$ and lattice cutoff
$\epsilon$. Let ${\rm arg}^\epsilon, \partial^\epsilon_\mu,
\Delta^\epsilon$ denote the angle--function, the derivative
in the $\mu-$direction and the laplacian in the $\epsilon-$ lattice,
respectively, then
using the equality
$$
\epsilon_{\mu\nu} \partial^\epsilon_\nu {\rm arg}^\epsilon (x-y) =
\partial^\epsilon_\mu (\Delta^\epsilon)^{-1} (x-y),
$$
we immediately obtain

\be
{1\over |\Lambda|} \int_\Lambda d^2x \bar V^2_z (x) = -{1\over |\Lambda|}
\sum_{\scriptstyle \ell,k\atop\scriptstyle{\underline{x}} \subset\Lambda}
q_\ell q_k (\Delta^\epsilon)^{-1} (x_\ell, x_k)
\simeq - {1\over |\Lambda|} \sum_{\scriptstyle \ell, k\atop{\underline{x}}
\subset\Lambda} {q_\ell q_k \over 2\pi} \ln (|x_\ell - x_k|+
\epsilon).
\label{lambda}
\ee

Equation (\ref{lambda}) looks like the energy per unit volume $E_{N,\Lambda}
({\underline{x}})$ of a neutral two--component system of $N$ particles with
charges $\pm \sqrt{1\over 2\pi}$ in a volume $|\Lambda|$ interacting via 2D
Coulomb potential with ultraviolet cutoff $\epsilon$. The average of
$E_{N,\Lambda} ({\underline{x}})$ over the positions of the charges
in the limit
$|\Lambda| \nearrow {\bf R}^2, N\nearrow \infty$ with fixed average density
$\delta$ can be identified with the free energy of the above system at
$\beta=1$ and chemical potential $\delta$ in the thermodynamic limit, and this
gives our estimate of $<\bar V^2_z>$.
The behaviour of the free energy can be obtained through a sine--Gordon
transformation\cite{edw} and for small $\delta$ it is given by\cite{fro76}:

\be
<\bar V^2_z> \sim -\delta \ln \delta.
\label{vz2}
\ee

Hence in our crude approximation $<\bar V^2_z>$ acts as a mass term
increasing with $\delta$. If we assume that the scaling limit and perturbation
expansion in  $<V^2_z>$  commute with each other, we find  that its effect
is to
drive the NL$\sigma$ model at large scale to the disordered massive
regime with a mass gap for $\vec\Omega$  of the  order of $m^2_s (\delta)
\sim - \delta
\ln \delta$.  Within this approximation, for  a more careful analysis one
should
consider the renormalization group equations for $g$ and $\lambda$
including the perturbation $<\bar V^2_z>$--term from the beginning. Then
the value of $m^2 (\delta)$ could be renormalized.

A slightly better approximation is to consider the holons as slowly moving
randomly distributed impurities, creating a random potential and analyse
the behaviour of the spin--wave $\vec\Omega$ in the presence of this
potential (this can again be justified for the limit $J/t <<1$ with a large
effective mass of holes). At a fixed time the random potential behaves as

$$
\bar V^2_z (w) \sim \sum_{{\underline{x}},{\underline{y}}} q_y q_x {(x -
w)^\mu \over |x - w|^2} {(y - w)^\mu \over |y - w|^2}.
$$
Hence it is positive and roughly falling like $r^{-2}$, where $r$ is the
distance to the closest impurity. This kind of systems have been considered
in \cite{esc} and it is expected to be in the localized regime for
$\vec\Omega$,
where (random) averaged Green functions  exhibit a mass gap $m(\delta)$
roughy proportional to the inverse mean free path. Hence up to logarithms
$m^2_s (\delta) \sim \delta$ for  small $\delta$.

\subsection{Crossover from Short Range to Long Range Antiferromagnetic Order}

All the above arguments strongly suggest that the spinon system described by
the action $S^\star_s$ in (\ref{spinac}) exhibits a mass gap
$m_s (\delta)$ increasing with
$\delta$ (at least for sufficiently large, but, nevertheless $\delta << 1$),
 vanishing at $\delta =0$, thus
showing an expected crossover to the insulating ``N\'{e}el state''
at zero doping\cite{deltac}. As
a result of $m(\delta)> 0$, the constraint $\vec\Omega^2 = f$ is relaxed at
large scales\cite{fra} and a lagrange multiplier field introduced in the
action
to impose the constraint would just mediate short--range attractive
interaction
 between spinons with opposite spin, with strength $0 (m_s
(\delta)^{-2} )$ rotationally symmetric in spin space (see \cite{fro2} for the
discussion of an analogous situation).
Another short--range interaction, but uniaxial in the spin space, is
introduced by the perturbation term $S'_s$ in (\ref{spinac}),
 with strength $0(m^2_s (\delta)/ g)$.

We can summarize the above discussion by rewriting the NL$\sigma$ model
effective action for spinons in the $CP^1$ form, neglecting short range
interactions, as

\be
S^\star_s = \int_{[0,\beta] \times {\bf R}^2} d^3x {1\over g}
\left[|(\partial_0 -
z^*_\beta
\partial_j z_\beta) z_\alpha |^2 + v^2_s |(\partial_\mu - z^*_\beta
\partial_\mu z_\beta) z_\alpha|^2 + m^2_s (\delta) z^*_\alpha z_\alpha\right].
\label{spinac1}
\ee

In the NL$\sigma$ model without mass term $(\delta=0)$ the constraint
$z^*_\alpha z_\alpha =f$ and the symmetry breaking condition, e.g. $<z_2>
\not = 0$, lead to excitations described by a complex massless field $S
\simeq  z_1 <z_2^*>$ with massless ``relativistic'' dispersion relations,
corresponding to spin waves.
In the NL$\sigma$ model with mass term the absence of symmetry breaking
and the effective softening of the constraint lead to excitations described
by the spin ${1\over 2}$ two--component field $z_\alpha$, with massive
``relativistic'' dispersion relations. However, the self--generated gauge
field $z^*_\alpha \partial_a z_\alpha, a = 0,1, 2$ would confine the spin
${1\over 2}$ degrees of freedom into composite spin 1 spin--wave fields.
On the other hand, as we shall see later, the coupling to the holons
will induce deconfinement of spin  ${1\over 2}$ excitations.

To give a kind of ``microscopic'' interpretation of the above results using
the slave
fermion picture in terms of the hard--core bosonic fields $b_\alpha,
b_\alpha^*$, we follow the approach of Yoshioka\cite{yos}.
First consider the case  $\delta =0;$ then  a MF treatment with an
$s$--like RVB
order parameter yields an energy gap vanishing at $({\pi\over 2}, {\pi\over
2})$, $(-{\pi\over 2}, -{\pi\over 2})$ in the Brillouin zone and an expansion
for low momenta around these 2 points shows that the corresponding
excitations are described by a complex massless field with relativistic
dispersion relations, which can be identified with a spin--wave field $S$;
the ground state is then the insulating N\'{e}el state \cite{yos}. A MF
treatment
with an $s+id$ RVB order parameter yields an energy gap vanishing at $(\pm
{\pi\over 2}, \pm {\pi\over 2})$ in the Brillouin zone, and an expansion for
low momenta around these 4 points shows  that the corresponding excitations
are described by a two--component massless field with relativistic
dispersion relation which can be identified with a spinon field
$z_\alpha$\cite{yos}.
These excitations are turned into massive ones in our approach by the $m_s
(\delta)$ term.
In some sense,  our result in the NL$\sigma$ model corresponds to a
kind of slave--fermion approach with a gauge--invariant RVB order
parameter $s+id$--like and minimal gap $m_s (\delta)$ at $(\pm {\pi\over 2},
\pm {\pi\over 2})$.  We should emphasize, however, that our approach is
different from the slave-fermion formalism\cite{dif} which considers long-range
ordered AF state with gapless excitations, while we consider short-range
ordered
AF state with gapful excitations.

	To the best of our knowledge, the present formulation is the first
successful attempt to explicitly include the AF fluctuations into the RVB-type
scheme of treating the strong correlation effects  in a self-consistent
manner. Upon doping the long range AF is destroyed, being replaced
by short-ranged AF order, which is physically obvious.
We have obtained an explicit doping dependence of the gap value which
has the correct extrapolation at zero doping (gap vanishes)\cite{dm}.
The doping dependence of the  AF correlation length $\xi \sim m^{-1} \sim
\delta^{-1/2}$, expected from our calculation, agrees very well with the
neutron
scattering data\cite{bir}.
Here we also propose a new interpretation of the spin gap effect
in underdoped superconductors -- mainly due to the short range AF order.
The recent numerical simulations\cite{st} seem to support our interpretation.

\section{ Holon effective action}
\label{holon}

\subsection{ Holon-Spinon Coupling via Self-Generated  Gauge Field}

Now  turn to holons. We use a gauge choice of the field $\bar B$ such
that $e^{i\int_{<ij>}\bar B}$ is purely imaginary and assume, following eq.
(\ref{hop}) -- (\ref{gtilde}) that the gauge -- invariant AM parameter is
$\simeq 1$,
which permits us to use the equality (see eqs. (\ref{spinac}), (\ref{rmat}))

\be
\langle \alpha_{<ij>} b^*_{i1} b_{j1} + \alpha^*_{<ij>} b_{i2} b_{j2}
\rangle \simeq 1 = b^*_{i1} b_{i1} + b^*_{i2} b_{i2}.
\label{bconst}
\ee

We obtain a low energy continuum effective action for holons by rescaling the
lattice spacing to $\epsilon <<1$ and neglecting higher order terms in
$\epsilon$.
Making use of (\ref{bconst}) one obtains

\bea
&
S_h = \dsp \int_0^\beta dx^0 \left\{  \sum_j H_j^* (\partial_0 -
(b^*_{j\alpha} \partial_0
b_{j\alpha})^{\#(j)} - \delta) H_j
+ \sum_{<ij>} (-) t e^{\tst i\int_{<ij>} \bar B} \Bigl\{ {H^*_i H_j - H^*_j
H_i \over
\epsilon} \right.&\cr
& \dsp \left. + (H^*_i H_j + H^*_j H_i) \Bigl[(b^*_{i\alpha} ({ b_{j\alpha}
- b_{i\alpha}
\over \epsilon})
- ({b^*_{i\alpha} - b^*_{j\alpha} \over \epsilon})
b_{j\alpha}\Bigr]^{\#(j)}\Bigr\}
+ O (\epsilon) \right\}, &
\label{holonac}
\eea
where $\#(j)$ denotes complex conjugation if $j$ is even. Neglecting the
$b$--terms, the action (\ref{holonac}) describes the usual two--component
Dirac
(``staggered'') fermions of the flux phase, with vertices of the double--cone
dispersion relations in the reduced Brillouin zone, centered (in the
$\epsilon=1$ lattice) at ($\pm {\pi\over 2}, \pm {\pi\over 2})$ and
chemical potential $\delta$. To define a continuum effective action, using the
standard  procedure  for the flux phase\cite{fra}, we first define 4
sublattices:
1) for $j_1, j_2$ even, (2) for $j_1$ odd, $j_2$ even, (3) for $j_1$ even
$j_2$ odd, (4) for $j_1, j_2$ odd. They can be grouped into two
``N\'{e}el'' sublattices
$A=\{(1), (4)\}$, $B=\{(2), (3) \}$. The holon field $H$ restricted to the
sublattice $\#$ is denoted by $H^{(\#)}$. The holon action can then be
written in the continuum limit as a bilinear form in $H^*$ and $H$ with kernel:

\be
\left(
\ba{cccc}
\partial_0 - z^*_\alpha \partial_0 z_\alpha -\delta,\; & \; i t
(\partial_1 +
z^*_\alpha \partial_1 z_\alpha), \;& \; - it (\partial_2 + z^*_\alpha
\partial_2 z_\alpha), \;& \;0
\cr
it (\partial_1 - z^*_\alpha \partial_1 z_\alpha), &\partial_0 + z^*_\alpha
\partial_0 z_\alpha -\delta, & 0. & i t (\partial_2 - z^*_\alpha \partial_2
z_\alpha)\cr
- it (\partial_2 - z^*_\alpha \partial_2 z_\alpha), & 0, & \partial_0 +
z^*_\alpha \partial_0 z_\alpha -\delta, &it (\partial_1 - z^*_\alpha
\partial_1 z_\alpha),\cr
0, & it (\partial_2 + z_\alpha^* \partial_2 z_\alpha), & it (\partial_1 +
z^*_\alpha
\partial_1 z_\alpha),& \partial_0 - z^*_\alpha \partial_0 z_\alpha -\delta
\ea
\right),
\label{matrix}
\ee
where we used the decomposition of $b_\alpha$ carried out in the previous
Section (all ferromagnetic terms are $0 (\epsilon)$ and so they do not
appear in the continuum limit (\ref{matrix})).

Setting

$$
\gamma_0 = \sigma_z, \qquad \gamma_\mu = (\sigma_y, \sigma_x), \qquad \A =
\gamma_\mu A_\mu,
$$
$$
\Psi_1 =\left(\matrix{\Psi^{(A)}_1 \cr
\Psi^{(B)}_1 \cr} \right)
= \left(\matrix{e^{-i{\pi\over 4}}
H^{(1)} + e^{i{\pi\over 4}}  H^{(4)} \cr
e^{-i{\pi\over 4}} H^{(3)} + e^{i{\pi\over 4}} H^{(2)} \cr} \right)
$$
$$
\Psi_2= \left(\matrix{\Psi^{(B)}_2 \cr
\Psi^{(A)}_2 \cr} \right)
= \left(\matrix{e^{-i{\pi\over 4}}
H^{(2)} + e^{i{\pi\over 4}}  H^{(3)} \cr
e^{-i{\pi\over 4}}  H^{(4)} + e^{i{\pi\over 4}}  H^{(1)} \cr} \right)
$$
\be
\bar \Psi_\# = \gamma_0 \Psi^\dagger_\#
\label{twocomp}
\ee
and assigning charge $e_A = +1, e_B = -1$ to the fields corresponding to
$A$
and $B$ sublattices, respectively, the continuum effective action
for holons can be rewritten as\cite{dm1}:

\be
S^\star_h = \int_{[0, \beta] \times {\bf R}^2} d^3x \sum^2_{r=1} \bar\Psi_r
\Bigl[\gamma_0 (\partial_0 - \delta - e_r z^*_\alpha \partial_0 z_\alpha) +
+ t (\parz - e_r z^*_\alpha \parz z_\alpha) \Bigr] \Psi_r
\label{holonac1}
\ee

Hence, $S^\star_h$ describes the coupling of the holon field to the
spinon--generated gauge field $z^*_\alpha \partial_a z_\alpha$ in a
``relativistic'' Dirac -- like form, with opposite coupling for the two
N\'{e}el sublattices. From (\ref{holonac1}) it is clear that (neglecting
the gauge terms),
the upper components of $\Psi_\#$ describe gapless excitations with small FS
($\epsilon_F \simeq 0 (\delta t))$, whereas the lower components describe
massive excitations with a doping dependent gap $m_h (\delta) \simeq
2\sqrt{\delta}$. In
terms of lattice fields we have excitations of the two bands (gapless and
gapful) supported in the reduced Brillouin zone due to the presence of the
N\'{e}el
sublattices. This leads to a band--mixing due to the non--diagonal
structure of the $\gamma$--matrices and this in turn yields a ``shadow band''
effect, i.e. a reduction of the spectral weight for the  outer part of the
F.S.
facing $(\pi, \pi)$, in agreement with experimental data\cite{mar} and MF
numerical simulation\cite{dai}. The FS of underdoped superconductors
has also been considered in the $SU(2)$ gauge field theory\cite{wen}.
However, there are some differences between the results obtained there
and the predictions of the present model (for details see \cite{dai}).

\subsection{$U(1) \times SU(2)$ Gauge Invariance of the Total Action}

Neglecting the  quartic fermion term, which would produce a short-range
repulsion between
holons with opposite charge, the full continuum effective action can be
rewritten introducing an auxiliary $U(1)$ gauge field $A$ (over which
one integrates in the path--integral):

\bea
&S^\star = \int_{[0,\beta] \times {\bf R}^2} d^3 x \left\{ {1\over g}
\Bigl[|(\partial_0 - A_0)
z_\alpha |^2 + v^2_s |(\partial_\mu - A_\mu) z_\alpha |^2 \right.& \cr
& \left. + m^2_s (\delta) z^*_\alpha z_\alpha] + \sum^2_{r=1} \bar\Psi_r
[\gamma_0
(\partial_0 - e_r A_0 -\delta)
+ t (\parz - e_r \A)]\Psi_r. \right\}
\label{totalac}
\eea

The action $S^\star$ is invariant under the $U(1)$-- gauge transformation

$$
\Psi_r (x) \rightarrow e^{\tst i e_r \Lambda (x)} \Psi (x),
$$
$$
\bar\Psi_r (x) \rightarrow e^{\tst -i e_r \Lambda (x)} \bar\Psi (x),
$$
$$
z_\alpha (x) \rightarrow e^{\tst i\Lambda(x)} z_\alpha (x), \quad
z^*_\alpha (x)
\rightarrow e^{\tst - i\Lambda (x)} z^*_\alpha (x),
$$
\be
A_a (x) \rightarrow A_a (x) + \partial_a \Lambda (x), \quad \Lambda (x) \in
{\bf R}.
\label{u1gauge}
\ee

Going backwards we recognize in (\ref{u1gauge}) the continuum limit of the
$h/s$ gauge
invariance (\ref{hsgauge}) which is gauge-fixed imposing, e.g., a Coulomb
gauge condition on
$A_a$.

\underbar{Remark} -- One easily finds that the lattice counterpart
of the $A_\mu$
gauge--fixing is a gauge--fixing for ${\rm arg} \ (\Sigma_i^* P e^{i
\int_{<ij>} V} \Sigma_j)$ which is $U(1)\times SU(2)$--gauge
invariant as required in Remark in Section II.

In fact

$$
\arg (\Sigma_i^* P e^{\tst i\int_{<ij>} V} \Sigma_j) \simeq \arg \
(\alpha_{<ij>}
b^*_{i1} b_{j1} + \alpha^*_{<ij>}
b^*_{i2} b_{j2}) \simeq z^*_\alpha \partial_\mu
z_\alpha \simeq A_\mu,
$$
where in the last equality we have neglected a quadratic fermion term
fictitiously
introduced in $S^\star$ (compare it  with (\ref{holonac1})).

We can use $S^\star$ to compute, as in\cite{il}, the gauge field propagator
induced
by the spinon and holon vacuum polarizations. In the Coulomb gauge,
denoting by
$\Pi^\bot (\Pi^\Vert)$  the transverse (longitudinal) polarizations,
respectively,  we have

$$
\langle A_\mu (\omega, q) A_\nu (-\omega, - q) \rangle =
\left(\delta_{\mu\nu} - {q_\mu q_\nu \over q^2} \right) {1\over (\Pi^\bot_s +
\Pi^\bot_h )(\omega, q)},
$$
$$
\langle A_0 (\omega, q) A_0 (-\omega, -q) \rangle = {\omega^2 \over
q^2(\Pi^\Vert_s + \Pi^\Vert_h) (\omega, q)}.
$$

Since the spinon is massive the vacuum polarization of the spinon system
alone would be Maxwell--like in the absence of holons and it would induce a
logarithmic confinement of spinons $(\Pi^\bot \simeq \omega^2 + q^2, \quad
\Pi^\Vert \simeq \omega^2$ for small $\omega, q$). However, due to their
gapless
component the holons induce a transverse vacuum polarization with Reizer
singularity\cite{ln} $(\Pi^\bot_h \simeq |{\omega \over q}| + \chi q^2)$ and a
leading behavior in $\Pi^\Vert_h \simeq {\omega^2\over q^2}$ for small
$\omega/q$, leading to a short range $A_0$ propagator, so that the full gauge
interaction is \underbar{not} confining\cite{am2}.

As a result, holons and spinons are true dynamical degrees of freedom in the
model. Nevertheless, since we are in 2D, the attractive force mediated by
the gauge field is expected to produce bound states with the quantum numbers
of the spin wave (see\cite{fro2} for the discussion of a similar problem) and
presumably (see e.g. \cite{chen}) of the electron.
This could explain why neglecting the (expected) semionic nature of holons
and spinons $(\delta V = 0 =\delta B)$ can be justified to some extent in
2D. On the contrary, in 1D the spinons and holons are deconfined, and their
statistical properties as semions are crucial for a proper account of
the spin-charge separation.

\section{ Conclusions}

To summarize our results it is interesting to compare the features
appearing in the (underdoped) 2D and 1D $t-J$ model (for small $\delta, J/t)$
within the $U(1) \times SU(2)$ gauge approach followed here and in \cite{marc}.

In both cases, applying to the $t-J$ model the $U(1) \times SU(2)$
C-S bosonization in terms of gauge fields $B, V$, we separate spin
and charge degrees of freedom of the electron, and describe them in terms of
spinon $(z)$ and holon ($\Psi$) fields.

1) The low--energy properties of the spinons can be described by a NL$
\sigma$-model: in 2D it is massive, and the mass gap
(vanishing at $\delta=0$) is due
to the coupling to holons mediated by a non--pure gauge $SU(2)$ field $\bar
V$; in 1D no such non--trivial gauge field exists and the spinons are
massless. The self--generated spinon--gauge field $(z^*_\alpha \partial_a
z_\alpha$) would confine the spinons into spin waves, but this is prevented in
2D by the coupling to holons, while in 1D by the presence  of a topological
$\theta= \pi$ term in the NL$\sigma$ model, which is absent in 2D.

2) The low--energy properties of holons are described in 2D by a Dirac
action, inducing band mixing, whose appearance is due to the presence of a
non--pure gauge field $\bar B$ (characteristic of the flux phase); the
absence of such a field in 1D implies that holon are described in term of a
spinless fermion action.

3) Spinons and holons in 2D are coupled by the spinon--generated gauge
field, carrying one degree of freedom (after gauge fixing) and this
presumably induces binding; the gauge fixing kills the degree of freedom of an
analoguous gauge field in 1D and as a conseguence, spinons and holons are
free.
The statistics is then dictated by the $U(1) \times SU(2)$ decomposition
and the corresponding (gauge--invariant) fields obey semionic statistics.

\bigskip

To reiterate, we outline some of the distinct features of our present study:

-- The short range AF order is the main reason of the existence of the
spinon gap (at least in the underdoped samples)
  which can explain a variety of experimental observations.

-- Neglecting gauge fluctuations
the 2--point correlation function for the electron exhibits half--pocket FS
in the reduced Brillouin zone, with a pseudogap minimal on the diagonal,
``shadow band'' effect and a quasi--particle peak due to the holons. These
features have been demonstrated in the MF computation of \cite{dai},
 where a similar action
for the system has been used, with a twist in holon--spinon statistics,
i.e., the spinons are fermions, while the holons are bosons.

-- The spin--wave persists even in the region without AF long range order,
 but as a short
ranged field.

-- We expect (this is still under investigation) that the ``composite
electron'' exhibits a Non--Fermi liquid behaviour due to the appearance of
Reizer singularity, thanks to transverse gauge fluctuations, in the
self--energy of its holon constituent.

	To conclude, the C-S bosonization approach, in spite of its technical
complications, is promising in providing a natural framework to describe
the normal state properties of underdoped superconductors.
After having settled in this paper our framework for a discussion of the
$t-J$ model at small $\delta$ and $J/t$ as a model for underdoped high $T_c$
cuprates, in the forthcoming paper  we will compute  correlation functions
and compare the results with experimental data\cite{marc1}.

\acknowledgments

We would like to thank J. Fr\"ohlich, Ch. Mudry, N. P. Ong,  F. Toigo and
 A. Tsvelik for
stimulating discussions. The work of P.M. was partially supported by
TMR Programme ERBFMRX-CT96-0045.

\appendix
\section{}

The proof of the bound (\ref{bound}) uses techniques adapted from the proof
of the
diamagnetic  inequality given in \cite{diam}.

The main ingredients are:

1) invariance of the action (\ref{auxiac}) under the gauge transformations:

$$
A_j \rightarrow A_j + \partial_0 \Lambda_j, \quad
U_{<ij>} \rightarrow U_{<ij>} e^{\tst i(\Lambda_i - \Lambda_j)},
$$
\be
H_j \rightarrow H_j e^{\tst i\Lambda_j}, \quad
H^*_j \rightarrow H^*_j e^{\tst -i\Lambda_j}, \quad \Lambda_j \in {\bf R}.
\label{gaugetr}
\ee

2) reflection (O.S.) positivity in the  absence of gauge fields: define an
antilinear involution $\theta$ on the holon fields supported in the
positive--time lattice (the time interval is identified with  $[-{\beta
\over 2}, {\beta \over 2}])$ by

$$
\theta: H_j (\tau) \rightarrow H^*_j (-\tau), \quad
H^*_j (\tau) \rightarrow H_j (-\tau)
$$
\be
\theta (AB) = \theta (B) \theta(A), \quad A, B \in {\cal J}_+
\ee
where  ${\cal J}_+$ denotes the set of functions of $H, H^*$ with support on
positive time. Reflection positivity is the following statement: $\forall F
\in {\cal J}_+$

\be
\langle F\theta F \rangle \equiv \int {\cal D} H {\cal D} H^* e^{-S(H,
H^*)} F\theta F^* \geq 0,
\label{ineq}
\ee
where $S(H, H^*) \equiv S(H, H^*, 0,1)$ (see eq. (\ref{auxiac})).

The involution $\theta$ defines an inner product ($\cdot, \cdot$) on ${\cal
J}_+$ by $\langle \cdot \theta \cdot \rangle$ and as a  consequence, Schwartz
inequality holds: $\forall F, G\in {\cal J}_+$

\be
|\langle F\theta G \rangle| \leq \langle F\theta F\rangle^{1\over2}
\langle G\theta G\rangle^{1\over 2}.
\label{sch}
\ee

Although the proof of (\ref{ineq}) is standard\cite{jaffe}, let us sketch
it for readers'
convenience and to set up a notation useful in the proof of (\ref{bound}).

We discretize the time interval introducing a time lattice with spacing
$\epsilon =\beta 2^{-N}$ and we choose the time$-0$ plane lying halfway
between lattice planes. We denote by $S^\epsilon_+, S^\epsilon_-$ and
$S^\epsilon_{+-}$ -- the terms of the discretized action existing in the
positive, negative time lattice and coupling the two, respectively, then we
have:

$$
S^\epsilon_- = \theta S^\epsilon_+
S^\epsilon_{+ -} = - \sum_j \bigl[ H^*_j ({\epsilon\over 2}) H_j ({-
\epsilon \over
2}) + H_j ({\epsilon \over 2}) H^*_j ({-\epsilon \over 2})\Bigr]
$$
\be
= - \sum_j \Bigl[ H^*_j ({\epsilon \over 2}) \theta H^*_j ({\epsilon \over
2}) + H_j
({\epsilon \over 2}) \theta H_j ({\epsilon \over 2}) \Bigr].
\label{refl}
\ee

Let $P_+$ denote the set of functions of $H, H^*$ given by linear
combinations with positive coefficients of elements of the form $F\theta
F$, $F\epsilon {\cal J}_+$. It is easy to see that $P_+$ is closed under
multiplication and summation with positive coefficients and for $\sum_i c_i F_i
\theta F_i \in P_+$ we have

$$
\int ({\cal D} H {\cal D} H^*)_\epsilon \sum_i c_i F_i \theta F_i =
\sum_i c_i |\int({\cal D} H {\cal D} H^*)_\epsilon F_i |^2 \geq 0,
$$
where $({\cal D} H {\cal D} H^*)_\epsilon$ denotes the (formal) measure on
$H, H^*$ in the path--integral for the discretized time model.

From (\ref{refl}) it follows that, for $F\in {\cal J}_+,
e^{-S^\epsilon (H, H^*)} F\theta F \in P_+,$ so  that

$$
\int ({\cal D} H {\cal D} H^*)_\epsilon e^{-S^\epsilon (H, H^*)} F\theta F
\geq 0
$$
and taking the limit $\epsilon \searrow 0$ we obtain (\ref{ineq}).

We turn  now to the proof of (\ref{bound}). We work again with
the discretized time lattice.
First  we use the gauge invariance (\ref{gaugetr}) to set $A=0$ on the
links in
$S_{+-}$. We denote by $\{A^{(1)}_1, U^{(1)}_1 \}$ $(\{A^{(1)}_2,
U^{(1)}_2\})$ the restriction of the gauge fields to the positive
(negative) time lattice and we set

$$
F_1^{(1)} = e^{\tst -[S^\epsilon_+ (H, H^*, A^{(1)}_1, U^{(1)}_1) -
S^\epsilon_+
(H, H^*)]}
$$
\be
F_2^{(1)} = e^{\tst -[S^\epsilon_+ (H, H^*, - r A^{(1)}_2, r U^*_2) -
S^\epsilon_+ (H, H^*)]},
\ee
where $r$ is reflection w.r.t. the time-0 plane.
Applying (\ref{sch}) we derive the upper bound:

$$
|\Xi^\epsilon (A, U) |\equiv |\int{\cal D} H {\cal D} H^* e^{\tst -S^\epsilon
(H, H^*, A, U)} |=
$$
$$
= |\langle F_1^{(1)} \theta F_2^{(1)} \rangle | \leq \langle F_1^{(1)}
\theta F_1^{(1)} \rangle^{1\over 2} \langle F^{(1)}_2 \theta F_2^{(1)}
\rangle^{1\over 2}.
$$

We  now choose a time plane, different from the time$-0$ plane, lying
halfway
between lattice planes, use the time translational invariance of $\Xi$
(inherited
from the antiperiodicity in time of $H, H^*$) to bring this plane to the
position of the time$-0$ plane shifting the whole lattice and repeat the
above procedure both for $F^{(1)}_1 \theta F^{(1)}_1$ and $F_2^{(1)} \theta
F_2$ (which are no more in the form $F\theta F$ with $\theta$ referred to
the new time$-0$ plane). As a result, we obtain, with obvious notations,
$F_\ell^{(2)}, \ell =1, 2^2,$ and then iterate. Finally  we derive the bound

\be
|\Xi^\epsilon (A, U))| \leq \prod^{2^N}_{\ell=1} \langle F_\ell^{(N)} \theta
F_\ell^{(N)} \rangle^{2^{-N}}.
\label{xibound}
\ee

A close inspection  shows that the gauge fields appearing in $F_\ell^{(N)}
\theta F_\ell^{(N)}$ have the following properties: $A=0$ and $U$ is time
independent. Using a hamiltonian transcription, with obvious meaning of
notations,  looking at (\ref{auxiac}), we improve the bound by means of
Golden--Thompson inequality:

$$
\langle F_\ell^{(N)} \theta F_\ell^{(N)} \rangle = Tr e^{\tst - \beta
\{[H_1 + H_2
(U=0)]+ [H_2 - H_2 (U=0)]\} } e^{\tst -\beta [H_2 - H_2 (U=0)]}
$$
\be
\leq Tr e^{\tst -\beta[H_1 + H_2 (U=0)]} e^{\tst -\beta[H_2 - H_2 (U=0)]}
\leq Tr e^{\tst -\beta[H_1 + H_2 (U=0)]},
\label{gt}
\ee
where we used $[H_2 - H_2 (U=0)] \geq 0$. Eq. (\ref{gt}) is equivalent
in the path-integral
formalism to bounding the r.h.s. of (\ref{xibound}) setting everywhere
$U=0$ in
$S^\epsilon_2$. Let us define

\be
\hat \Xi^\epsilon (U) \equiv \int ({\cal D} H {\cal D} H^*)_\epsilon
e^{\tst -S^\epsilon_1 (H, H^*, 0, U) + S^\epsilon_2 (H, H^*, 0)}|_{\partial_0
U=0}
\ee
and let $\hat U$ denote the gauge field configuration maximizing
$\hat \Xi^\epsilon (U)$, then the r.h.s. of (\ref{xibound}), using
(\ref{gt}), is bounded by
$\hat \Xi^\epsilon (\hat U)$ and taking the limit $\epsilon \searrow 0$ we
recover (\ref{bound}).


\begin{references}
\bibitem{deltac}   To be more precise, the  AF long range order (AFLRO)
does not
disappear at infinitesimal dopings, and there is a finite $\delta_c$ below
which
the AFLRO survives.  However, in our approach we have not considered in
detail the mechanism by which the AFLRO disappears  (we have not carried out
explicitly the renormalization procedure of the NL$\sigma$ model at
finite dopings, etc.).  Since we are mainly interested in the region where
only short range AF order exists, this should be acceptable. To make a
``partial
remedy'', probably, it is more correct to replace $\delta$ by $\delta -
\delta_c$
in some of our formulas.
\bibitem{dif} First of all, the standard slave-fermion formalism involves only
the $U(1)$ gauge field, whereas our model deals with both $U(1)$ and $SU(2)$
gauge fields. Secondly, there is an obvious difference
with the slave-fermion approach, namely,
the contraint in our case is: $\sum_\alpha b^*_\alpha b_\alpha =1$, while
for the latter it is $\sum_\alpha b^*_\alpha b_\alpha + h^* h=1$, with $h$
as the holon operator. Moreover, these two approaches are still different
 after a "renormalization" to take care of the above difference, as explained
in the main text.
\bibitem{am2}  The normal state properties of cuprates
were also studied earlier using the C-S approach\cite{am1}. However, the gauge
field propagator in that formulation is still fully relativistic, so the
plausible
non-Fermi liquid behavior considered there has a different origin.
\end{references}
\end{document}